\documentclass[prx,twocolumn,superscriptaddress,nofootinbib,noshowpacs,preprintnumbers,longbibliography,floatfix]{revtex4-2}
\usepackage[utf8]{inputenc}
\usepackage{graphicx}
\usepackage{float}
\usepackage{amssymb}
\usepackage{amsmath}  
\usepackage{adjustbox}
\usepackage{mathtools}
\usepackage{dsfont}
\usepackage{array}
\usepackage{bm,fixmath}
\usepackage{mathrsfs}
\usepackage{pifont}
\usepackage{multirow}
\usepackage{upgreek}
\usepackage{xcolor}
\usepackage{bm}
\usepackage{bbm}
\usepackage{physics}
\usepackage{comment}
\usepackage{tikz}
\usepackage[normalem]{ulem}
\usepackage{qcircuit}
\usepackage{color, soul}
\usepackage{placeins}
\usepackage{multirow}
\usepackage{booktabs}
\usepackage[pdftex,
            pdftitle={Resource-Efficient Simulations of Particle Scattering on a Digital Quantum Computer},
            pdfauthor={Yahui Chai, Joe Gibbs, Vincent R. Pascuzzi, Zoe Holmes, Stefan Kuehn, Francesco Tacchino, Ivano Tavernelli},
            bookmarks,
            colorlinks,
            linkcolor=myblue,
            citecolor=mymagenta,
            menucolor=black,
            urlcolor=myblue,
            plainpages=false,
            pdfpagelabels,
            hypertexnames=false]{hyperref}
\usepackage{orcidlink}

\graphicspath{{figures/}}

\definecolor{mymagenta}{RGB}{200, 0, 100}
\definecolor{myblue}{RGB}{45, 48, 146}
\definecolor{mypurple}{RGB}{200, 112, 255}

\newcolumntype{C}[1]{>{\centering\arraybackslash}p{#1}}

\begin{document}

\title{Resource-Efficient Simulations of Particle Scattering on a Digital Quantum Computer}

\author{Yahui Chai}
\thanks{These authors contributed equally to this work.}
\affiliation{Deutsches Elektronen-Synchrotron DESY, Platanenallee 6, 15738 Zeuthen, Germany}

\author{Joe Gibbs}
\thanks{These authors contributed equally to this work.}
\affiliation{School of Mathematics and Physics, University of Surrey, Guildford, GU2 7XH, UK}
\affiliation{AWE, Aldermaston, Reading, RG7 4PR, UK}

\author{Vincent R. Pascuzzi}
\email{vrpascuzzi@ibm.com}
\affiliation{IBM Quantum, IBM Thomas J.\ Watson Research Center, Yorktown Heights, NY 10598, USA}

\author{Zo\"{e} Holmes}
\affiliation{Institute of Physics, Ecole Polytechnique F\'{e}d\'{e}rale de Lausanne (EPFL),   Lausanne, Switzerland}
\affiliation{Centre for Quantum Science and Engineering, Ecole Polytechnique F\'{e}d\'{e}rale de Lausanne (EPFL),   Lausanne, Switzerland}
\affiliation{Algorithmiq Ltd, Kanavakatu 3 C, FI-00160 Helsinki, Finland}

\author{Stefan Kühn}
\affiliation{Deutsches Elektronen-Synchrotron DESY, Platanenallee 6, 15738 Zeuthen, Germany}

\author{Francesco Tacchino}
\affiliation{IBM Quantum, IBM Research Europe - Zurich, 8803 Rueschlikon, Switzerland}

\author{Ivano Tavernelli}
\email{ita@zurich.ibm.com}
\affiliation{IBM Quantum, IBM Research Europe - Zurich, 8803 Rueschlikon, Switzerland}

\date{\today}

\begin{abstract}
    We develop and demonstrate methods for simulating the scattering of particle wave packets in the interacting Thirring model on digital quantum computers, with hardware implementations on up to 80 qubits. We identify low-entanglement time slices of the scattering dynamics and exploit their efficient representation by tensor networks. Circuit compression based on matrix product state techniques yields on average a reduction by a factor of $3.2$ in circuit depth compared to conventional approaches, allowing longer evolution times to be evaluated with higher fidelity on contemporary quantum processors. Utilizing zero-noise extrapolation in combination with Pauli twirling, on quantum hardware we accurately simulate the full scattering dynamics on 40 qubits, and further demonstrate the wave packet state-preparation on 80 qubits.
\end{abstract}

\maketitle

\section{Introduction}\label{sec:Introduction}
Scattering experiments are at the heart of unraveling the internal structure of matter and the interactions between the fundamental particles. Major experimental facilities such as LHC~\cite{LHC_Bruning:2012zz} and RHIC~\cite{RHIC_Harrison:2002es} continue to generate valuable experimental data to test the theoretical predictions and drive the search for new physics. 
Concomitantly, there are significant efforts to develop analytical and numerical methods for improving our understanding of gauge field theories, which provide the theoretical framework for particle physics. 

In particular, Lattice Field Theory (LFT) provides a powerful tool for exploring non-perturbative regimes from first principles. Discretizing a theory on a Euclidean space-time lattice allows for applying sophisticated Monte Carlo (MC) methods, that have been extremely successful for computing properties such as mass spectra, phase diagrams and many other static properties~\cite{Durr2008, Alexandrou_2014}. However, the conventional MC approach to LFT is not suited to directly explore dynamical problems, as it crucially relies on the formulation in Euclidean space-time, and using Minkowski space-time would lead to a sign problem preventing efficient MC sampling. While indirect approaches exist to address scattering problems with Monte Carlo methods, such as extracting scattering phase shifts using Lüscher's method~\cite{Luscher:1986pf, Luscher:1990ux}, these techniques become increasingly challenging to apply at high-energies or for inelastic scattering processes~\cite{Brice_o_2018, Hansen_2019, Davoudi_2021}. In addition, such indirect approaches do not provide detailed access to the real-time dynamics of the particles during or immediately after a collision. Hence, there is an ongoing effort to find alternative methods allowing for overcoming these limitations.

\begin{figure*}[t!]
    \centering
    \includegraphics[width=1.0\linewidth]{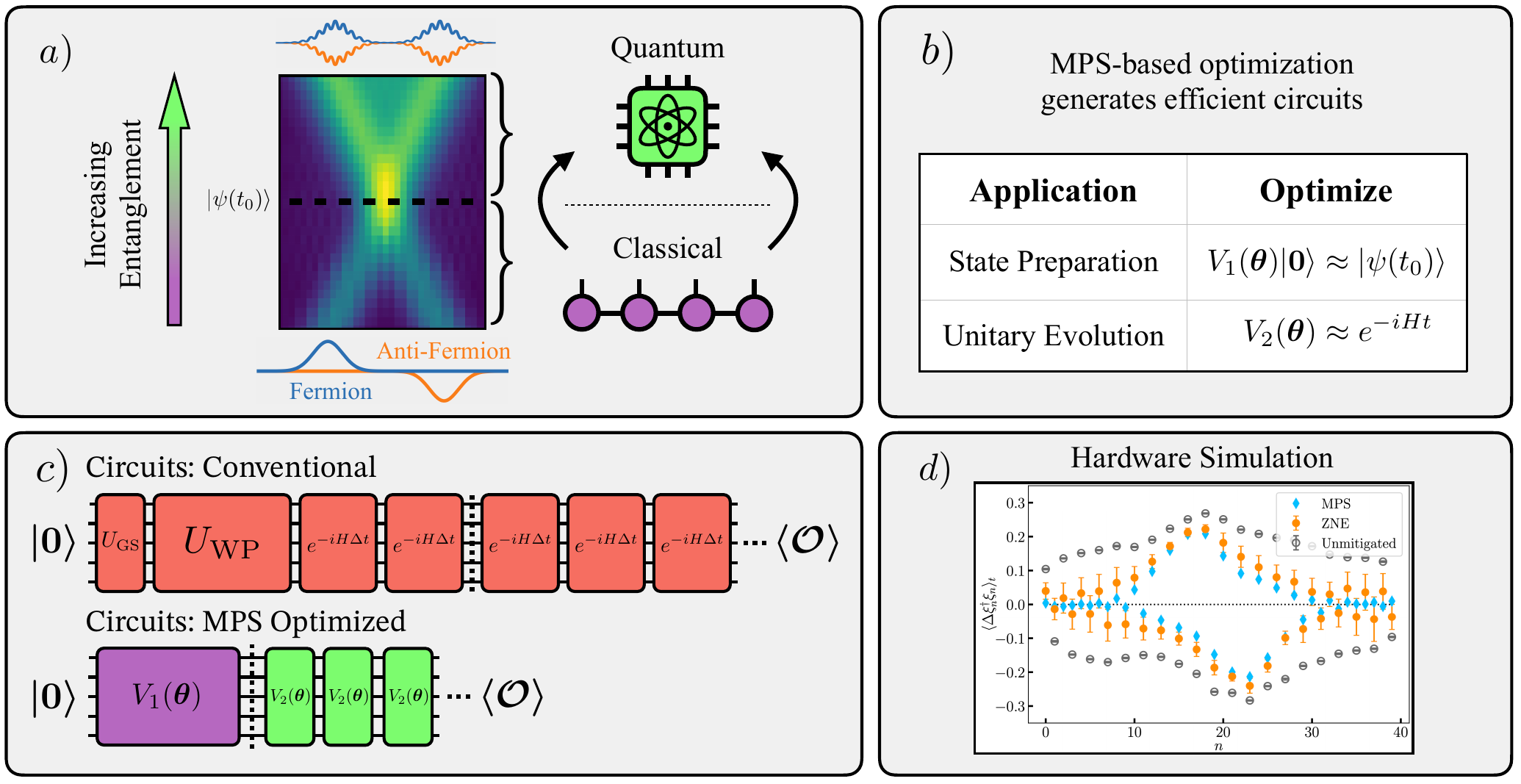}
    \caption{\textbf{Overview.}
    a) The target quantum simulation is a scattering interaction. Initially spatially separated wave packets only show entanglement (indicated by the brightness of the color) within each wave packet. As they move towards each other during the Hamiltonian evolution and interact, entanglement between the wave packets is generated. This may persist in long-range as they separate, motivating the use of the quantum computer to represent this highly entangled state, despite the weakly entangled initial wave packet configuration. To aid the simulation on near-term quantum hardware, we exploit the low-entanglement early dynamics, by simulating these times classically with tensor networks up to their limit, then the state is passed over to the quantum computer for further simulation.
    b) The hybrid use of tensor networks and the quantum computer is enabled by MPS-based circuit optimization, to find efficient and short-depth circuits for performing the simulation. This involves generating two circuits: $V_1(\boldsymbol{\theta})$ performs the initial state preparation of $|\psi(t_0)\rangle$, which has been pre-computed as an MPS; to continue the time evolution, we exploit the low-entangling property of $e^{-iH t}$ for small $t$, to learn a circuit $V_2(\boldsymbol{\theta})$ with a shorter depth than available Trotterizations.
    c) These circuits realize the same simulation as alternate conventional methods with shorter depth circuits, and we find up to a total factor of $3.2$ reduction, significantly improving the amenability to near-term quantum hardware.
    d) The combination of tensor network circuit optimization, performant quantum hardware, and modern error mitigation tools, enables a high-fidelity simulation of the full scattering dynamics on 40 qubits.
    }
    \label{fig:Overview}
\end{figure*}

An alternative approach for the simulation of scattering processes are tensor network techniques~\cite{orus2019tensor}. Numerical algorithms based on tensor networks take advantage of the classical simulability of slightly-entangled states~\cite{Vidal2003Efficient}, allowing for an efficient representation of the wave function in terms of low-rank tensors. In addition, numerical algorithms based on tensor networks do not suffer from the sign problem and have been demonstrated to allow for reliable calculations in regimes which are inaccessible with conventional MC methods~\cite{Banuls2017,Banuls2017a,Buyens2017,Kawauchi2018,Funcke2019,Silvi2017,Magnifico2021,Nakayama2022,Funcke2023,Itou2024}. In particular, Refs.~\cite{Pichler2016, Rigobello2021, Papaefstathiou2024, Belyansky2024} simulated meson scattering within the Schwinger model in both weak and strong coupling regimes using Matrix Product States (MPS). However, the substantial growth of entanglement following the collision in certain parameter regimes necessitates considerable computational resources to achieve accurate results. This poses a significant challenge for tensor networks, a difficulty that can become even more pronounced when trying to extend the approach to higher-dimensional settings~\cite{Pavesic2025}.

Quantum computing facilitates the simulation of real-time dynamics by constructing quantum circuits that approximate the time-evolution operator, e.g., through the use of Trotterization~\cite{Trotter1959, Suzuki1991,miessen2023quantum}. Several proposals have been made to extract asymptotic scattering observables from such simulations~\cite{turro2024, Erik2021, Erik2021_2}. Quantum computers also have the potential to simulate the full dynamics of the scattering process, allowing for capturing intermediate snapshots at any stage of the evolution~\cite{di2024quantum}. A central challenge is the preparation of the initial state, which should represent separated particle wave packets in position space with well-defined momenta. Recent works have addressed this and demonstrated simulations and first implementations on quantum processors~\cite{Chai2025fermionicwavepacket, Davoudi2024, Roland_2024, zemlevskiy2024, farrell2025, davoudi2025, chai2025, Julian_2025, joshi2025_scatteringqudit}. However, the simulation of dynamics up to long time-scales generally requires deep quantum circuits, making it difficult to obtain accurate results on current and near-future quantum hardware due to the levels of noise. This necessitates the development of novel methods that reduce circuit depth, while preserving fidelity, to fully leverage the potential of current quantum devices.

In this work, we seek to make the most of the combined strengths of quantum and classical simulation by leveraging the observation that tensor network methods can simulate the initial state and the dynamics of fermionic scattering processes at short times. In particular, we employ tensor network methods to variationally optimize quantum circuits~\cite{berezutskii2025tensor}, enabling an automated search for shallow-depth unitary blocks that are more compatible with current quantum devices. An increasing body of work has explored the use of tensor network methods to support quantum simulations across a range of applications via variational circuit compilation. These efforts include ground state preparations~\cite{dborin2022matrix, rudolph2023decomposition, rogerson2024quantum}, short-time state evolution~\cite{anselme2024combining, lin2021real, robertson2023approximate}, and the construction of short-depth circuits that approximate the unitary evolution operator $e^{-iHt}$ governed by a Hamiltonian $H$~\cite{causer2024scalable, gibbs2024deep, mc2023classically, mc2024towards, le2025riemannian, gibbs2025learning}.

Here we demonstrate our scalable tensor network compression strategy on the interacting Thirring model. That is, rather
than performing the full real-time evolution on hardware, we MPS simulations to compute the early-time dynamics and identify low-entanglement states. These states are then variationally compiled into short-depth circuits and used as starting points for hardware execution. To simulate the remaining real-time evolution, we apply circuit compression to the time-evolution operator as well. Specifically, we approximate short-time evolution segments $e^{-iHt}$ using variational circuits, where the target operator can be expressed as a matrix product operator (MPO) with high accuracy for small $t$. The variational circuit is optimized to maximize the overlap with the target time-evolution unitary, represented by this MPO. Compared to standard Trotterization, the resulting circuits are shallower in depth, enabling efficient execution of the full scattering process on noisy quantum hardware. Figure~\ref{fig:Overview} provides an overview of our approach.

The remainder of this paper is organized as follows. In Sec.~\ref{sec:Model}, we briefly review the lattice Thirring model and describe the construction of fermionic wave packets used to simulate scattering processes. The use of MPS pre-computation is described in Sec.~\ref{sec:CircuitOpt}, and the method used for variational circuit optimization for state preparation and time evolution is further detailed in Appendix~\ref{sec:circuit_opt_detail}. We motivate the scalability of our MPS-based approach with circuit compilations on up to 160 qubits.
Section~\ref{sec:results} presents our results demonstrating the techniques on quantum hardware. We find our methods enable the target simulation to be realized with a $3.2$ times reduction in circuit depth compared to the conventional approach, allowing the full scattering dynamics on 40 qubits to be simulated on quantum hardware with high-accuracy. We further give a hardware demonstration of the state preparation of the initial wave packets on 80 qubits.
Finally, in Sec.~\ref{sec:summary_outlook} we summarize our findings and discuss possible directions for extending this approach to more complex scattering processes and higher-dimensional models.

\section{Lattice formulation of the Thirring model\label{sec:Model}}
In this work, we use the Thirring model~\cite{Thirring1958} to study fermion scattering. While it can be solved exactly in the massless case using bosonization, and its spectrum can be determined using Bethe-ansatz in the massive case, it shares many interesting features with more complicated gauge field theories from the Standard Model. In particular, the Thirring model is renormalizable and can show scale-dependent behavior reminiscent of asymptotic freedom~\cite{Mari_Carmen}. Moreover, it exhibits behavior analog to confinement, where fermionic excitations cannot propagate freely.  Hence, it can serve as a testbed for new lattice techniques.

Adopting the Kogut-Susskind staggered formulation~\cite{Kogut1975, Susskind1977}, the lattice Hamiltonian of the model reads~\cite{Mari_Carmen}
\begin{equation}
    \begin{aligned}
    H &= \sum_{n}\left(\frac{i}{2a} \left( \xi_{n+1}^{\dagger}\xi_{n} - \xi_n^{\dagger}\xi_{n+1}\right) + (-1)^n m \ \xi_n^{\dagger}\xi_n \right)\\
    &+ \sum_{n} \frac{g(\lambda)}{a} \xi_n^{\dagger}\xi_n \xi_{n+1}^{\dagger} \xi_{n+1} \, ,    
    \end{aligned}
    \label{eq:hamiltonian_lattice_fermionic}
\end{equation}
where $\xi^{\dagger}_n$ and $\xi_n$ are fermion creation and annihilation operators; $a$ is the lattice spacing, $m$ is the fermion mass, and $g$ is coupling strengths of the four-fermion interaction term. Without loss of generality, we set $a=1$ for the rest of this work. 

In the non-interacting case ($g=0$) and with periodic boundary conditions, the Hamiltonian in Eq.~\eqref{eq:hamiltonian_lattice_fermionic} can be diagonalized in momentum space, $H=\sum_k w_k\left(c_k^\dagger c_k - d_kd_k^\dagger\right)$, where the momentum dependent operators are given by
\begin{equation}
    \begin{aligned}
        c_k^{\dagger} &= \frac{1}{\sqrt{N}} \sqrt{\frac{m+w_k}{w_k}} \sum_n e^{ikn} \left( \Uppi_{n0} + v_k \Uppi_{n1}\right) \xi_n^{\dagger}, \\
        d_k^{\dagger} &= \frac{1}{\sqrt{N}} \sqrt{\frac{m+w_k}{w_k}} \sum_n e^{ikn} \left( \Uppi_{n1} + v_k \Uppi_{n0}\right) \xi_n \, ,
    \end{aligned}
    \label{eq:operator_transformation}
\end{equation}
with $k \in {2\pi/N}\times\{ -\lfloor N/4 \rfloor, \cdots \lceil N/4 \rceil-1\ \}$ and $N$ the number of sites. The constants $v_k$ and $w_k$ correspond to 
\begin{equation} 
   v_k = \frac{\sin(k)}{m+w_k},\ \  w_k = \sqrt{m^2 + \sin^2(k)},
   \label{eq:wk_vk}
\end{equation}
and $\Uppi_{n0} ~(\Uppi_{n1})$ are projection operators defined as
\begin{align} 
    \Uppi_{nl} = \left(1+\left(-1\right)^{n+l}\right)/2,\quad\quad l\in\{0,1\} \, .
    \label{eq:projector}
\end{align}
The operators $c_k^{\dagger}$ ($d_k^{\dagger}$) create (annihilate) fermions in the position space, and are therefore referred to as fermion (anti-fermion) operators. 

Building on the work of Ref.~\cite{Chai2025fermionicwavepacket}, we investigate the scattering between a fermion and an anti-fermion wave packet for the interacting case. As shown in the reference, such wave packets can be created on top of the ground state $\ket{\Omega}$, by acting on it with a set of creation operators 
\begin{equation}\label{eq:initial_state}
    \ket{\psi_0} = D^{\dagger} C^{\dagger} \ket{\Omega}.
\end{equation}
Here $C^{\dagger} $  and $D^{\dagger}$ create, respectively, a Gaussian fermion and anti-fermion wave packet. These operators can be expressed as the linear combinations
 \begin{equation}
    \begin{aligned}
        C^{\dagger}(\phi^{c}) &= \sum_k \phi_k^c c_k^{\dagger} = \sum_n \tilde{\phi}_n^c \xi^{\dagger}_n, \\
        D^{\dagger}(\phi^d) &= \sum_k \phi_k^d d_k^{\dagger} = \sum_n \tilde{\phi}_n^d \xi_n,
    \end{aligned}
    \label{eq:creation_operators_gaussian_particle}
\end{equation}
where the Gaussian coefficients $\phi_k^{c(d)}$ in momentum space are given by
\begin{equation}
    \phi_k^{c(d)} =  \frac{1}{\sqrt{\mathcal{N}_k^{c(d)}}} e^{-ik\mu_n^{c(d)}} e^{-(k-\mu_k^{c(d)})^2 / 4\sigma_k^2}\, .
    \label{eq:coeff_momentum}
\end{equation}
In the expression above $\mu_n^{c(d)}$ corresponds to the position around which the wave packet is centered, $\mu_k^{c(d)}$ to the mean momentum, $\sigma_k$ represents the width in momentum space, and $\sqrt{\mathcal{N}_k^{c(d)}}$ is a normalization factor. 

Throughout our work, we set $\{\mu_k^{c}, \mu_k^{d}\} = \{4\times {2\pi}/{N}, -4\times {2\pi}/{N}\}$, $\sigma_k = {2\pi}/{N}$, and $\{\mu_n^{c}, \mu_n^{d}\} = \{N/4, 3N/4 - 1\}$.
The factors $\tilde{\phi}_n^{c(d)}$ in Eq.~\eqref{eq:creation_operators_gaussian_particle} represent the coefficients in position space, which can be obtained by Fourier transformation from the ones in momentum space as detailed in Eq.~(13) of Ref.~\cite{Chai2025fermionicwavepacket}. Although the operators defined above yield exact fermion (anti-fermion) wave packets only in the noninteracting limit under periodic boundary conditions, they nevertheless provide as a good approximation in the interacting regime, provided that the coupling constant $g$ is sufficiently small~\cite{Chai2025fermionicwavepacket}. 

To simulate the fermionic degrees of freedom on a quantum computer, we map them to Pauli matrices using the Jordan-Wigner transformation 
\begin{equation}
    \begin{aligned}
        \xi_n^{\dagger} &= \prod_{l<n} \ \sigma_l^z \sigma_n^{-}, \quad
        \xi_n  &= \prod_{l<n} \ \sigma_l^z \sigma_n^{+} \, , 
    \end{aligned}
    \label{eq:jw}
\end{equation}
where $\sigma^\pm_l =\left(\sigma^x_l\pm i\sigma^y_l\right)/2$ and $\sigma^j_l$, $j \in \{x, y, z\}$ are the usual Pauli matrices. The resulting qubit Hamiltonian in terms of Pauli operators then reads
\begin{equation}
    \begin{aligned}
    	H &= \frac{i}{2} \sum_{n=0}^{N-2} \left( \sigma_{n+1}^- \sigma_n^+ - \sigma_n^- \sigma_{n+1}^+ \right)  \\
    	  &+ \frac{m}{2} \sum_{n=0}^{N-1} (-1)^n \left( \mathds{1} - \sigma_n^z  \right) \\
          &+ \frac{g}{4}\sum_{n=0}^{N-2} \left( \mathds{1} - \sigma_n^z  \right) \left( \mathds{1} - \sigma_{n+1}^z  \right),
    \end{aligned}
    \label{eq:Hamiltonian_Paulis}
\end{equation}
where we have chosen open boundary conditions for simplicity. For the spatially localized wave packets we use in our simulations, boundary effects remain negligible, provided the system size is sufficiently large. In both our classical simulations and quantum hardware runs with $N=40$, no significant boundary effects were observed.

As pointed out in Ref.~\cite{Chai2025fermionicwavepacket}, simulating the scattering process generally involves the following steps. First, the ground state of the model has to be prepared. It can be obtained, for example, by a variationally optimized circuit $\ket{\Omega} = U_{\mathrm{GS}}\ket{0}$. Second, particle wave packets can be created on top of the vacuum by applying suitable operators as in Eq.~\eqref{eq:initial_state}. A corresponding quantum circuit for these operators can be constructed based upon Givens rotations (see Appendix~\ref{app: resources}, Fig.~\ref{fig:conventional_circuit}) to prepare this state $\ket{\psi_0} = U_{\mathrm{WP}}\ket{\Omega}$, serving as the initial state of the scattering process. Finally, the dynamics can then be realized by performing real-time evolution of the initial state up to a time $t$. This can be done using standard Trotterization, e.g. approximating the time evolution operator by a series of unitaries implementing a small timestep $\Delta t$ which can be efficiently realized on quantum hardware, $\ket{\psi(t)} = U(\Delta t)^{t/\Delta t} |\psi_0\rangle$. 

However, the quantum resource requirements for simulating a complete scattering process using the above protocol remain quite substantial. For a 40-qubit system, the total circuit depth exceeds 300, with over 5000 two-qubit gates (see Table.~\ref{tab:resource_compare} in Sec.~\ref{sec: circuit_reduction}). While a total number of around 5000 CNOTs may be feasible in practice, the primary limitation arises from the circuit depth and the associated two-qubit gate error and qubit decoherence. The details of this resource estimation can be found at Appendix~\ref {app: resources}. 

\begin{figure}
    \centering
    \includegraphics[width=\linewidth]{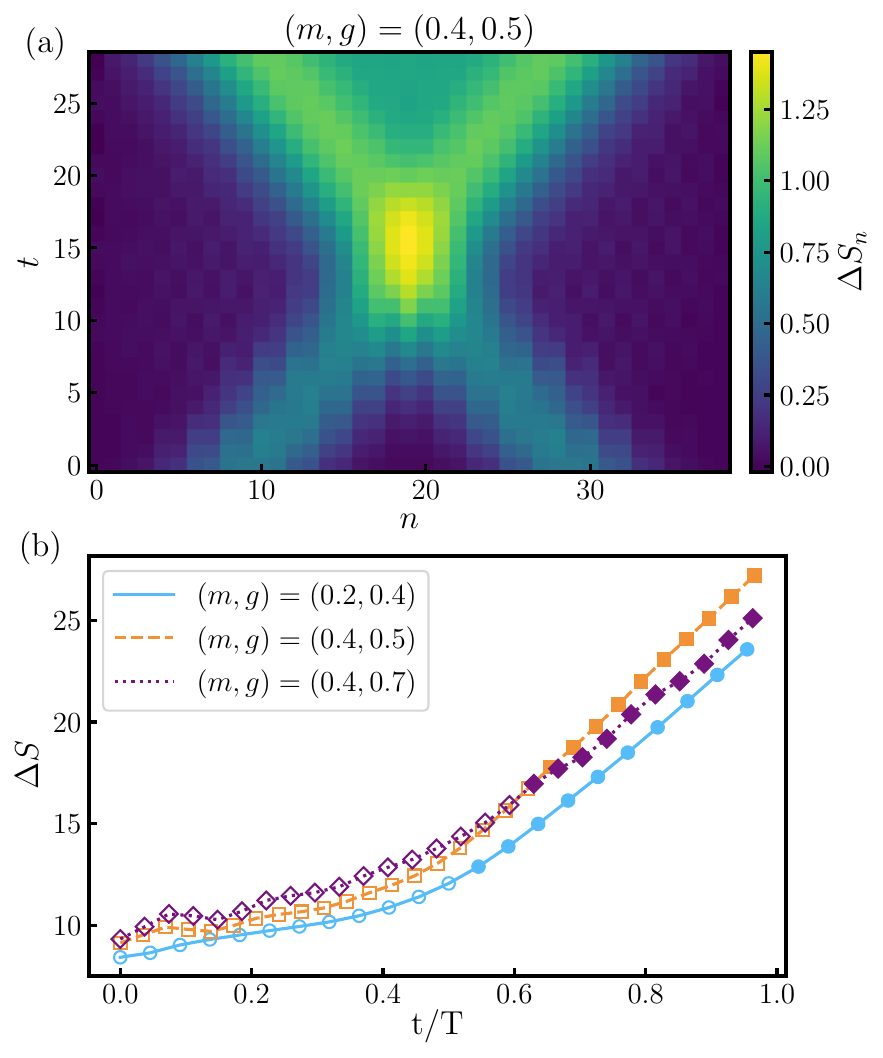}
    \caption{Entanglement entropy over time in the fermion-antifermion scattering process. (a) The bipartite entanglement entropy for the case $(m,g)=(0.4, 0.5)$ for $H$ in Eq.~\eqref{eq:Hamiltonian_Paulis}. The $x$-axis is the site index, and the $y$-axis represents time. (b) The sum of bipartite entanglement entropy across all nontrivial cuts, versus normalized time $t/T$. Empty markers represent the time slices simulated by tensor networks and compiled into the state preparation circuit, and the filled markers represent time slices that will be simulated directly on quantum hardware. The time $t$ is normalized with the total time $T$ to visualize different processes in the same plot, $T \in \{22, 29, 27\}$ for $(m, g) \in \{(0.2, 0.4), (0.4, 0.5), (0.4, 0.7)\}$ respectively. All simulations are performed using MPS with an SVD truncation threshold of $10^{-8}$. }
    \label{fig: entropy}
\end{figure}

Tensor networks can efficiently simulate low-entanglement dynamics. In the scattering process, entanglement typically increases after the collision, making the tensor network simulation more costly. However,  before the collision, especially at the early stage, entanglement remains relatively low, allowing for efficient and accurate tensor network simulation. To illustrate this, we take the fermion-antifermion scattering as an example and calculate the entanglement entropy of bipartition. Specifically, we calculate the von Neumann entropy $S_n(t) = -\tr[ \rho_n(t) \log_2 \rho_n(t) ]$, where $\rho_n(t)$ is the reduced density matrix of
the first $n$ qubits at time $t$. To quantify the entropy generated throughout the scattering process, we subtract the vacuum contribution and obtain the excess entropy $\Delta S_n(t)$. 

As shown in Fig.~\ref{fig: entropy}(a), before the collision, entanglement is mainly localized within each wave packets. At the collision point, where the wave packets overlap, entanglement peaks at the central sites. After the collision, significant entanglement is generated between the outgoing particles, and long-range entanglement persists as they propagate, leading to increasing simulation costs for tensor networks. 

To quantify the overall simulation complexity, Fig.~\ref{fig: entropy}(b) displays the total excess entanglement entropy $\Delta S(t) = \sum_n \Delta S_n (t)$. The figure shows $\Delta S$ continually increases over time, with a notable acceleration after the collision (marked by filled symbols). This behavior motivates a hybrid strategy: simulate the low-entangled states before collision by tensor networks, and continue the subsequent dynamics on quantum computers.  To realize this hybrid approach, we employ a recently developed circuit optimization algorithm based on tensor networks~\cite{gibbs2024deep}, which we apply to both state preparation and time-evolution circuits, as demonstrated in Fig.~\ref{fig:Overview}. This approach significantly reduces circuit complexity, enabling high-accuracy hardware execution for 40-qubit systems.

\section {Tensor Network Circuit Approximation}\label{sec:CircuitOpt}

To fully exploit the capabilities of present-day quantum hardware, it is essential to design circuits with compact depth.
To aid the generation of such circuits, we identify low-entanglement time slices of the dynamics that permit efficient representation by tensor networks. This enables a circuit optimization based on MPS, to find shorter depth circuits to realize the simulation than would be otherwise possible without the help of classical computation. Crucially, to go beyond the limits of tensor networks, these circuits are concatenated on the quantum hardware to explore states with a larger amount of entanglement.

In this direction, there are two natural circuits to produce using this optimization. 
The first circuit prepares the initial state followed by a period of time evolution,  $e^{-iHt}|\psi_0\rangle$. The initial state, $|\psi_0\rangle$, consisting of the two spatially separated wave packets on top of the vacuum, lacks long-range entanglement and therefore can be efficiently represented by an MPS. As the time evolution proceeds and the wave packets move together and interact, entanglement increases and a correspondingly higher bond dimension is required to accurately describe this state. Therefore we can use a tensor network time evolution algorithm to evolve the state $e^{-iHt}|\psi_0\rangle$ up to the largest time $t_0$ that still permits an accurate representation with a computationally tractable bond dimension. 
The second circuit exploits the low-entangling nature of $e^{-iHt}$ for small values of $t$. When the Hamiltonian, $H$, has a quasi-1D topology with local interactions, then $e^{-iHt}$ can be efficiently represented by a low bond dimension MPO.
Again, this permits a variational optimization to search for circuits approximating this unitary, with shallower circuits compared to those derived from Trotterization.

\subsection{Overview of Variational Circuit Approximation}\label{sec: variation_state}

Here we briefly describe the approach taken for the variational circuit approximation, with more details found in Appendix~\ref{sec:circuit_opt_detail} and Ref.~\cite{gibbs2024deep}. 

We focus on minimizing 2-qubit gate count and depth, as these dominate errors on near-term hardware. During the optimization, all variational circuits are constrained to have a 1D brickwork structure of layers of $SU(4)$ gates (the most general expression of a 2-qubit gate as a dimension-4 unitary matrix) with nearest-neighbor connectivity, resulting in layers of gates acting on `odd' and `even' pairs of qubits (as sketched in Fig.~\ref{fig: gate_env} in Appendix~\ref{sec:circuit_opt_detail}). All $SU(4)$ gates can be decomposed into at most three 2-qubit gates (c.f.\ Appendix~\ref{app: resources}, Fig.~\ref{eq: givens_fermion_op}), therefore a circuit composed of $l$ $SU(4)$ layers can be described by at most $3l$ layers of primitive 2-qubit gates (e.g. CNOT, CZ) when mapped onto the device native gate set. These layers of 2-qubit gates all act on distinct qubits, and therefore can be applied in parallel.

All our circuit optimizations can be separated into two categories. First, the generation of a short-depth circuit $V_1(\boldsymbol{\theta})$ to approximately prepare a target state 
$$V_1(\boldsymbol{\theta})|0\rangle \approx |\psi_\text{Targ}\rangle,$$ where $|\psi_\text{Targ}\rangle$ is represented by a low-bond dimension MPS. A successful optimization will output a circuit to a target fidelity with a lower gate count/depth compared to the alternate `conventional' approach for wave packet preparation described in Ref.~\cite{Chai2025fermionicwavepacket}. This motivates the following cost function for a numerical optimizer to minimize, measuring the infidelity between the target and variational state 
\begin{equation}\label{eq:C_State}
    C_{\text{State}}(\boldsymbol{\theta}) = 1 - \big| \langle \psi_{\text{Targ}}|V_1(\boldsymbol{\theta})|0\rangle \big|^{2} .
\end{equation}

Our second category of circuits involve simulating real-time evolution under the Hamiltonian.
$e^{-iHt}$, for small $t$, can be represented by a low-bond dimension MPO, and is computed using high accuracy MPO Trotterization.
We aim to use the variational optimization to produce a short depth circuit (\textit{shorter} than a standard Trotter depth circuit) to simulate the time evolution, such that
\begin{align}
    V_2(\boldsymbol{\theta}) \approx e^{-iHt}.
    \label{eq:time_evolution_op_approx}
\end{align}
This could be done by variationally searching for a shallower circuit to simulate a given Trotter step with a fixed error. Alternatively, given a fixed depth ansatz for the circuit, we can optimize the ansatz to minimize the error. We here take the latter approach.

This motivates the following cost function, measuring a Hilbert-Schmidt inner product between the unitaries
\begin{equation}\label{eq:C_Uni}
    C_{\text{Uni}}(\boldsymbol{\theta}) = 1 - \frac{1}{2^{2N}} \left|\text{Tr} (V_2(\boldsymbol{\theta})^\dagger e^{-iHt})\right|^2.
\end{equation}

Both circuits use the same optimization procedure, where the $SU(4)$ gates defining the circuits are iteratively updated by computing the gate's environment tensor, after which a Polar Decomposition is applied to find the optimal gate update. This a standard practice in tensor network optimization~\cite{evenbly2009algorithms}, which has been similarly motivated for circuit optimization~\cite{shirakawa2024automatic, rudolph2023decomposition, gibbs2024deep, causer2024scalable}. Further details on the tensor network based optimization can be found in Appendix~\ref{sec:circuit_opt_detail}.

\subsection{Initial State}\label{sec:InitialState}

In this section, we detail how the initial state wave-packet configuration at $t=0$ is generated as an MPS. First, the ground state of the Hamiltonian (see Eq.~\eqref{eq:Hamiltonian_Paulis}) in the half-filling $U(1)$ symmetry sector is computed as an MPS using the 2-site density matrix renormalization group (DMRG) algorithm~\cite{schollwock2011density}; we use the quantum number preserving MPS functionality provided by the \texttt{ITensorMPS} package~\cite{fishman2022itensor} to perform this calculation. For all Hamiltonian settings considered, the ground state can be represented by an MPS with a maximum bond dimension of 32, after retaining a minimum singular value of $10^{-12}$.

The ground state MPS is then used to create the state $|\psi_0\rangle$ describing the initial configuration of the separated fermion and antifermion wave packets. This state is exactly constructed using  Eqs.~\hyperref[eq:jw]{(\ref*{eq:initial_state}--\ref*{eq:jw})} with MPS arithmetic. The addition of MPS in this fashion substantially increases the bond dimension compared to the ground state. To find a more compact MPS representation, we perform a variational compression~\cite{schollwock2011density}. Namely, we sweep through the site tensors of a lower bond dimension variational MPS, maximizing the fidelity with the uncompressed MPS. We perform this compression for increasing bond dimension MPS, finding the smallest with infidelity less than $10^{-6}$ with respect to the uncompressed state, which resulted in an MPS with a maximum bond dimension $\chi \approx 20$.
\begin{figure}[t]
    \centering    \includegraphics[width=\linewidth]{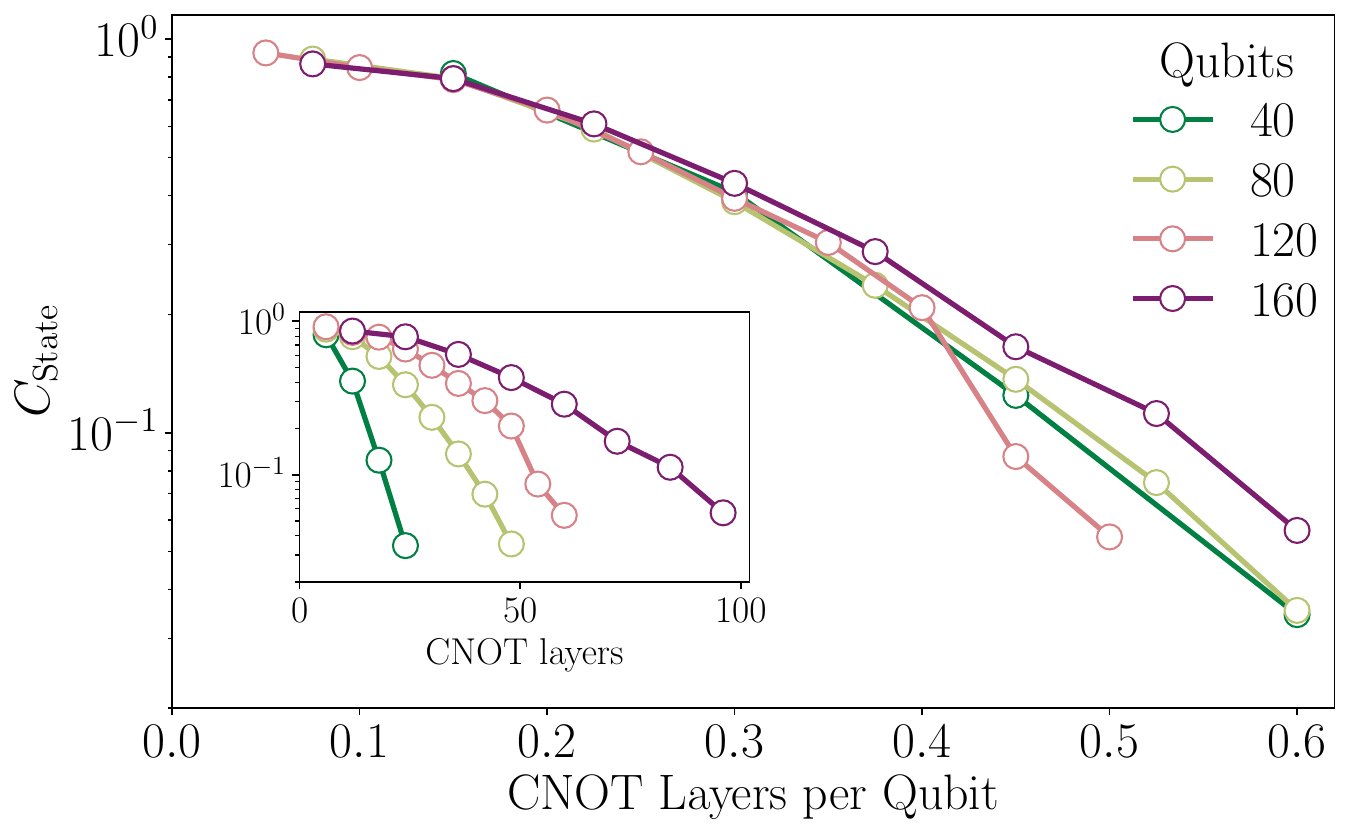}
    \caption{ \textbf{Scaling of Optimization - Initial State:} The variational optimization is performed to generate circuits to prepare the state $|\psi_\text{Targ}\rangle = |\psi_0\rangle$, representing the initial separated wave packets, on increasing size systems, with $N\in\{40,80,120,160\}$ qubits shown by the colored data. Each data point corresponds to the outputted infidelity cost function value after convergence of the optimization. When the depths of the circuits are scaled by the system size, as shown in the main plot where the $x$-axis plots the number of CNOT layers per qubit, a similar scaling across all qubit numbers is observed in achievable infidelity versus CNOT layers per qubit. The unscaled data (with the $x$-axis simply the number of CNOT layers) is shown in the inset.
    }
    \label{fig: tn_circuit_mps_vary_n}
\end{figure}

To probe the scalability of MPS-based circuit optimization we first compile circuits for preparing the initial wave packet configurations $|\psi_0\rangle$ with the Hamiltonian coefficients $(m,g)=(0.2,0.4)$ on increasing system sizes $N\in\{40,80,120,160\}$.
We perform the variational optimization for increasing depth circuits, with the value of the infidelity cost function $(C_\text{State})$ upon convergence plotted as a function of circuit depth, shown in Fig.~\ref{fig: tn_circuit_mps_vary_n}. When the depth is scaled with respect to the number of qubits in the system, a similar scaling is observed in the achievable infidelity to this target state versus the CNOT layers per qubit in the circuit. The quality of these approximate initial state preparations is further visualized in Fig.~\ref{fig:fermion_densities_scale_n} in the Appendix~\ref{app:fermion_densities_system_size}, by computing the fermion densities of the initial state. These state preparation results on up to 160 qubits provide strong evidence that MPS circuit optimization can aid quantum simulation on system sizes far beyond the limits of exact state vector simulation.
\begin{figure}[tp!]
    \centering
    \includegraphics[width=\linewidth]{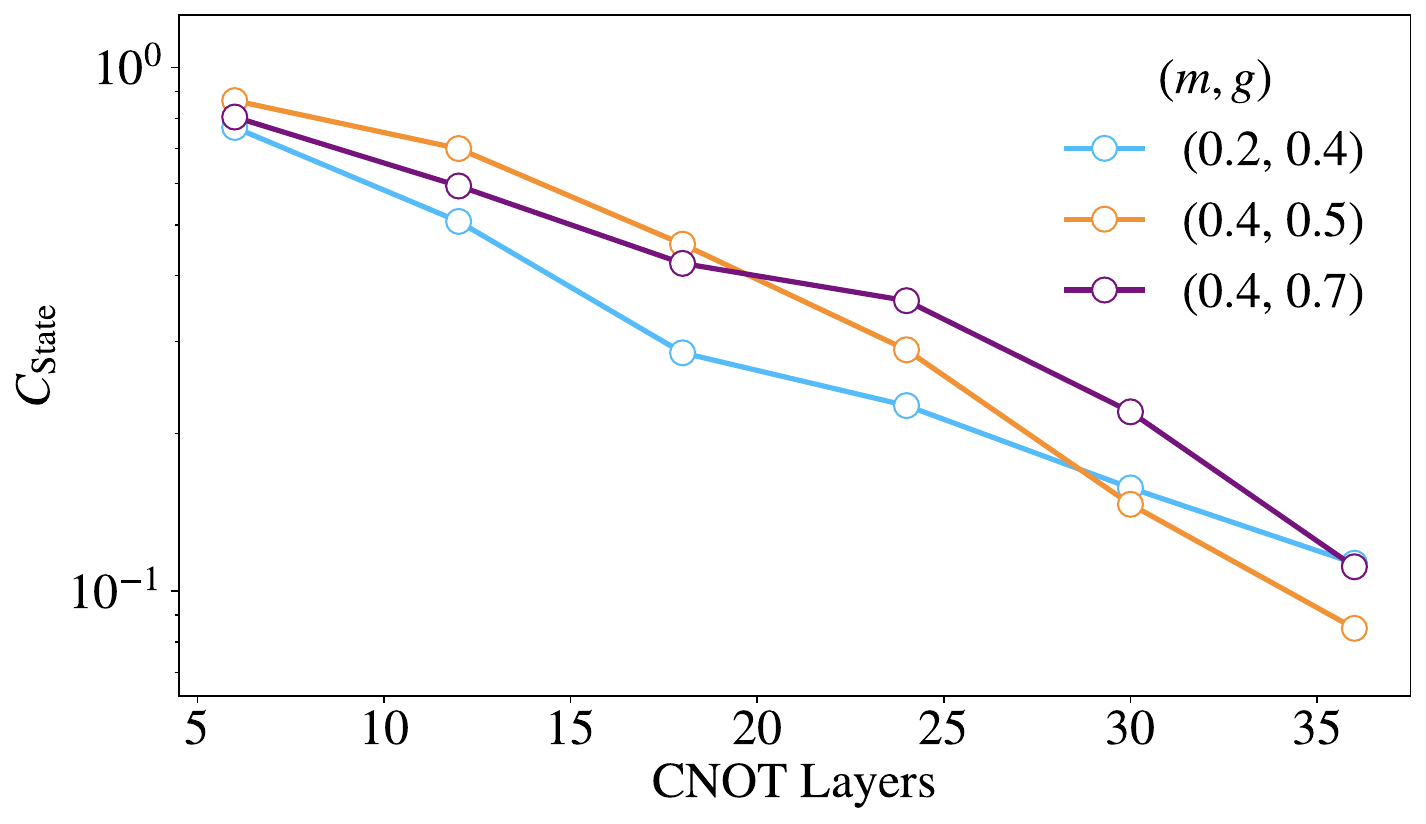}
    \caption{ \textbf{Optimization - Time Evolved Initial State:} Data showing final cost function value upon convergence of variational MPS optimization for generating the state $|\psi_\text{Targ}\rangle = e^{-iHt_0}|\psi_0\rangle$. The depth of the variational circuit measured in layers of commuting CNOT gates, the $y$-axis measures the corresponding value of infidelity ($C_{\text{State}}$) upon convergence of the optimization. The target state is computed using MPS arithmetic, followed by TEBD.     The 3 data curves correspond to the Hamiltonian settings considered, $(m,g) \in \{(0.2,0.4), (0.4,0.5), (0.4,0.7)\}$, with corresponding initial evolution times of $t_0 \in \{11,18, 16\}$ respectively. 
    }
    \label{fig: tn_circuit_mps}
\end{figure}

For our hardware demonstration, we further push the use of MPS pre-computation by incorporating some early time evolution steps into the target state. Concretely, we simulate the evolution up to where the wave packets begin to interact and residual entanglement entropy begins to be generated (as sketched in Fig.~\ref{fig:Overview}a)). We perform this time evolution using the time evolution block decimation (TEBD) method~\cite{vidal2004efficient} with a maximum MPS bond dimension of $\chi=150$, using a 2nd-order Trotterization of the propagator with a timestep of $\Delta t=0.25$. This computation results in an MPS representing the target state $|\psi_\text{Targ}\rangle = e^{-iHt_0}|\psi_0\rangle$. The variational optimization is performed three times to find circuits to approximately prepare the target state, for Hamiltonians with coefficients $(m,g) \in \{(0.2,0.4), (0.4,0.5), (0.4,0.7)\}$ and corresponding initial evolution times $t_0 \in \{11,18, 16\}$. The result of these optimizations is presented in Fig.~\ref{fig: tn_circuit_mps}. 
For $(m,g)=(0.2,0.4)$, a circuit depth of 30 CNOT layers is used for $V_1(\boldsymbol{\theta})$, and for $(m,g) \in \{(0.4,0.5), (0.4,0.7) \}$ a circuit depth of 36 CNOT layers is used.

\subsection{Time Evolution Unitary}\label{sec: variation_timeevolution}

To continue the dynamics, we further perform real-time evolution under the Hamiltonian Eq.~\eqref{eq:Hamiltonian_Paulis} on the quantum computer. This requires the implementation of the unitary $e^{-iHt}$. To derive a circuit to approximate this unitary, the standard approach would be to employ a Trotterization. Under this approach, there is always a trade-off between the Trotter error versus the circuit depth. We further take advantage of tensor network circuit optimization to search for both short-depth and low error circuits to approximate this unitary.

Given the Trotter errors we can tolerate, and the low depth overhead compared to 1st order Trotter for 1D nearest neighbor Hamiltonians, the 2nd-order Trotterization gives the shortest depth circuits out of the available Trotter orders. Therefore the circuit $V_2(\boldsymbol{\theta})$ is initialized with the equivalent depth 2nd-order Trotterization, and any reduction in error versus $e^{-iHt}$ is an advantage (i.e., has a lower error than the equivalent-depth Trotter step circuit).

\begin{figure}[htp!]
    \centering
    \includegraphics[width=\linewidth]{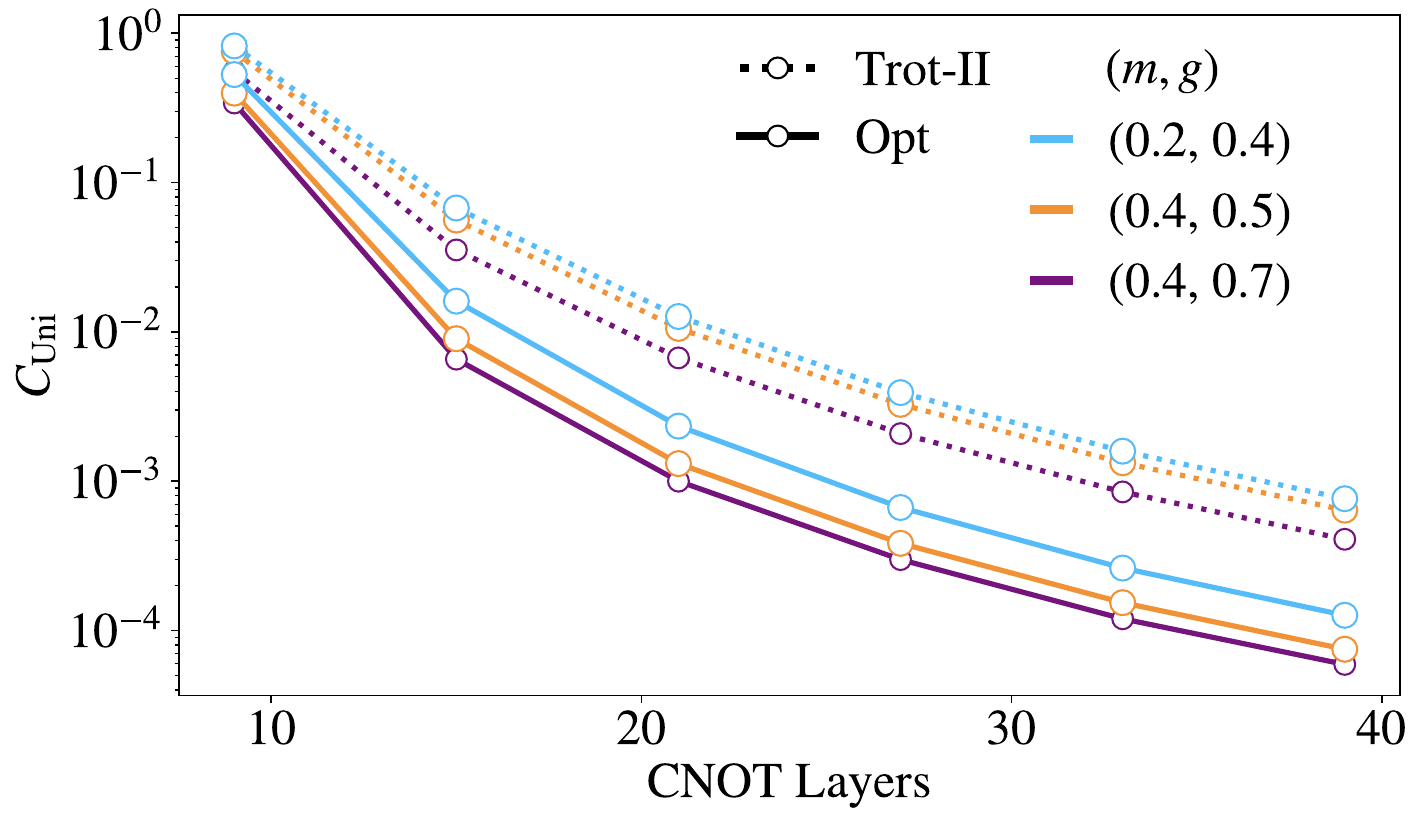}
    \caption{\textbf{Optimization - Time Evolution Unitary:} Data comparing the approximation error to the target unitary $e^{-iH t}$, with timestep $t = 2.0$, for increasing depth circuits. The data connected by the dotted line labeled `Trot-II' corresponds to increasing depth 2nd-order Trotterizations of the target unitary. The full line labeled `Opt' displays the approximation error, $C_\text{Uni}$,  of increasing depth circuits derived from the MPO-based optimization. 
    }
    \label{fig: tn_circuit_mpo}
\end{figure}

The MPO-based optimization updates the $SU(4)$ gates via Polar Decomposition of gate environments to minimize the cost function Eq.~\eqref{eq:C_Uni}, as described in Ref.~\cite{gibbs2024deep}. Figure~\ref{fig: tn_circuit_mpo} shows the results of this optimization. 
The time evolution unitary, $e^{-iHt}$ with $t = 2.0$, is computed with high-accuracy Trotterization converged in timestep, represented as an MPO with bond dimension $\chi=128$. For each of the three Hamiltonian parameter pairs, $(m,g) \in \{(0.2,0.4), (0.4,0.5), (0.4,0.7)\}$, we find that (excluding the shortest depth circuit of 9 CNOT layers) the optimization is able to reduce the error $C_{\text{Uni}}$ compared to the equivalent depth 2nd-order Trotterization by 6.60, 8.00 and 5.52 respectively on average across the range of circuit depths tested. For all hardware demonstrations we perform the time evolution using the optimized circuits with a circuit depth of 15 CNOT layers. We highlight this has a lower error than the second-order Trotterization circuits with a depth of 21 CNOT layers, an over 25\% reduction in circuit depth.

\subsection{Overall Circuit Depth Reduction}\label{sec: circuit_reduction}

We highlight the overall reduction in circuit depth for full simulation of scattering dynamics gained through the use of the tensor network optimization, in comparison to the alternate construction previously described in Ref.~\cite{Chai2025fermionicwavepacket}. Here, the circuit suggested can be summarized by $U(\Delta t)^{t/\Delta t} U_{\text{WP}} U_{\text{GS}}|0\rangle$, where $U_{\text{GS}}$ prepares the vacuum state $|\Omega\rangle$, $U_{\text{WP}}$ is an operator preparing the initial wave packet configuration on top of the vacuum, and $U(\Delta t)$ is the Trotter unitary used to simulate real-time evolution. We give circuit depth estimates for these three circuits separately.

The depth of $U_\text{GS}$ to prepare $|\Omega\rangle$ is negligible compared to the total circuit depth, so its contribution is omitted. For an $N$-qubit system, the CNOT depth of $U_\text{WP}$ decomposed by Givens rotations is around $2N-4$. Similarly, the CNOT depth due to time evolution by second-order Trotterization over total time $T$ is given by $6T / \Delta t + 3$, see Appendix~\ref{app: resources}
for further details. The sum of these two formulas give a lower-bound on the total CNOT depth 
\begin{equation}\label{eq: D_Conv}
    D_{\text{Conv}} = 2N+6T/\Delta t - 1.
\end{equation}
In our hardware experiments of the full scattering process we implement the simulation on $N=40$ qubits. We observed in Fig.~\ref{fig: tn_circuit_mpo} that an error of $C_\text{Uni} \approx 0.01$ for approximating the time-evolution circuits provided results that were sufficiently consistent with high-accuracy MPS simulation. A second-order Trotterization depth with the same (Trotter) error would use $\Delta t = 2/3$. The total evolution time is chosen such that at the end of the simulation the scattered wave packets have fully spread to the edge of the boundaries. For the three Hamiltonian settings $(m,g) \in \{(0.2,0.4), (0.4,0.5), (0.4,0.7)\}$, we find (via classical simulation) that this corresponds to $T \in \{21, 28, 26\}$. Therefore, we can finally compute the corresponding depths for the same simulation using the conventional approach would be $D_\text{Conv} =$ 268, 331, 313.

In comparison, the simulation using our MPS-optimized circuits for Hamiltonian settings $(m,g) \in \{(0.2,0.4), (0.4,0.5), (0.4,0.7)\}$ required total circuit depths of $D_\text{Opt} = 90, 96, 96$ respectively. By computing the relative reduction in circuit depth, given by $D_\text{Conv} / D_\text{Opt} = 2.98, 3.44, 3.26$, we find that on average the total circuit depth of the simulation has been reduced by a factor of $3.23$ compared to the conventional approach, a significant reduction which greatly improves the ability to perform these simulations on state-of-the-art noisy processors. 
For our 80 qubit hardware demonstration of state preparation, the tensor network optimized circuits can realize the target state $e^{-iHt_0}|\psi_0\rangle$ with a significantly reduced circuit depth of 25 CNOT layers, down from 249 if using the conventional approach, a depth reduction by a factor of 10.0.
Table~\ref{tab:resource_compare} summarizes these improvements.
\begin{table}[htp!]
    \centering
    \begin{tabular}{@{} lcc @{}}
        \toprule
        \multicolumn{3}{c}{$N = 40$, $T = 28$, $\Delta t = 2/3$} \\
        \midrule
            & CNOT layers & CNOT gates \\
        \midrule
        $\ket{\psi(t_0)}$    & 36 (241)   & 702 (3371)   \\
        \midrule
        $e^{-i2H}$ & 12 (18)  & 234 (351)  \\
        \midrule
        In total       & 96 (331)  & 1872 (5126)  \\
        \bottomrule
        \toprule
        \multicolumn{3}{c}{$N = 80$, $t_0 = 10$, $\Delta t = 2/3$} \\
        \midrule
            & CNOT layers & CNOT gates \\
        \midrule
        $\ket{\psi(t_0)}$    & 25 (249)   & 948 (3987)   \\
        \bottomrule

    \end{tabular}
    \caption{Comparison of two-qubit gate depth in terms of distinct CNOT layers and total CNOT gate number using the tensor network optimized circuit versus the conventional approach (in brackets). \newline
    \textit{Top} - Resources for 40 qubit hardware simulation in Sec.~\ref{sec: hardware_40}. For $(m,g) = (0.4,0.5)$ the state before collision $\ket{\psi(t_0 = 18)}$ can be prepared with 36 CNOT layers using the tensor network optimized approach as shown in Sec.~\ref{sec: variation_state}. In comparison, the conventional way, requires 241 CNOT layers to prepare $\ket{\psi(t_0=18)}$ as estimated in Eq.~\eqref{eq: D_Conv}. To realize the unitary $e^{-i2H}$ to sufficiently low error, the MPS optimized circuits required 12 CNOT layers while the second-order Trotterization required 18. Trivial layer compression in the Trotterization (due to neighboring layers of gates acting on identical pairs of qubits) is included. \newline
    \textit{Bottom} - Resources for 80 qubit hardware simulation in Sec.~\ref{sec: hardware_80}. Here, the circuit depth reduction compared to the conventional approach is even greater, with a reduction by a factor of 10.0.
    }
    \label{tab:resource_compare}
\end{table}

\section{Benchmarking on quantum hardware\label{sec:results}}
\begin{figure}[htp!]
    \centering
    \includegraphics[width=0.8\linewidth]{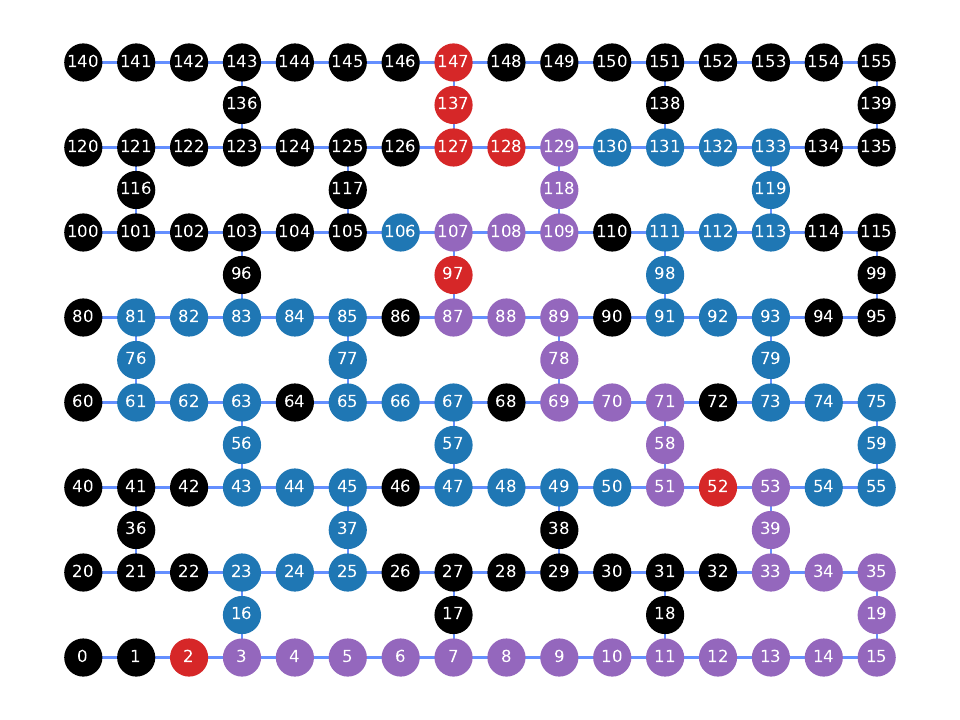}
    \caption{\textbf{Layout of \texttt{ibm\_fez}}. The circles indicate the 156 qubits, lines connecting circles indicate qubit pairs between which a CZ gate can be directly implemented. For all of our experiments, we choose a linear chain of qubits on the heavy-hex lattice depending on the current error rates of the chip. The qubits highlighted in red indicate the linear chain used for the experiment with $N=40$ and $(m,g) = (0.4, 0.7)$ at $t = 26$. The qubits highlighted in blue are those used for the $N=80$ initial state preparation experiment with $(m,g) = (0.2,0.4)$. The qubits highlighted in violet were used in both experiments.
    }
    \label{fig:fez-layout}
\end{figure}

We execute the resulting tensor network optimized circuits on an IBM quantum computer for different parameter sets to demonstrate that our approach can accurately capture different physical behaviors.
The experiments were run on IBM Quantum's \texttt{ibm\_fez}, a Heron r2 processor with 156 fixed-frequency transmon qubits with tunable couplers on a heavy-hex lattice layout (see Fig.~\ref{fig:fez-layout}). Here, we present device properties at the time of experimentation. The mean readout error was 1.2\% and the median was 1.0\%. A higher mean is indicative of an asymmetry that leads to higher qubit readout errors skewing the overall distribution of errors. The relaxation time, $T_1$, had a mean of $149~\mu s$ and a median of $140~\mu \text{s}$, and dephasing times, $T_2$, with mean and median times of $103~\mu \text{s}$ and $98~\mu \text{s}$, respectively.
The single-qubit gates had a mean error rate of $1.4 \times 10^{-4}$ and median of $6.4 \times 10^{-5}$. Lastly, the two-qubit gate errors had a mean of $3.7 \times 10^{-3}$ and a median of $3.4 \times 10^{-3}$. In our experiments, a linear chain of qubits was chosen on the heavy-hex lattice trying to exclude individual qubits with low $T_1$ and $T_2$ times as well as large error rates (c.f.\ Fig.~\ref{fig:fez-layout}).

\subsection{Simulating dynamics for 40-qubits}\label{sec: hardware_40}
In this subsection, we demonstrate hardware runs of the scattering process for 40-qubit systems. Three representative cases are considered: $(m,g) \in\{(0.2,0.4), (0.4,0.5),(0.4,0.7)\}$, corresponding respectively to cases where the fermion and antifermion pass through each other, exhibit partially repulsive interaction and experience strong repulsion. In all cases, we chose the state $\ket{\psi(t_0)}$ shortly before the collision as the starting state for the hardware run, corresponding to $t_0 \in \{11, 18, 16\}$, as mentioned in Sec.~\ref{sec: variation_state}. These states have relatively low entanglement compared to the states after collision, and can be approximated by shallow, trainable quantum circuits, as outlined in the previous section. The subsequent timesteps are executed on the quantum hardware, by applying the compressed quantum circuit for evolving the system for $t=2$ to the starting state (see Table~\ref{tab:resource_compare} for the exact gate numbers). 

The Qiskit \texttt{Estimator} primitive is used to evaluate expectation values along the dynamics. This class provides methods to perform an array of error suppression and mitigation techniques in a unified way, making it straightforward to estimate expectation values of observables. We configure an \texttt{Estimator} with dynamical decoupling (DD) using the XY4 sequence to suppress decoherence on idle qubits, while Twirled Readout Error Extinction (TREX)~\cite{Berg_2022} is used to mitigate readout errors. In addition, Pauli twirling~\cite{Wallman_2016} is applied to two-qubit gates to transform coherent errors into stochastic Pauli noise, allowing for more robust extrapolation to the zero-noise limit when estimating physical observables. For each circuit, we generate 200 random twirled instances and perform 1000 measurements per instance to estimate the expectation values of Pauli-$Z$ operators. To further mitigate errors, we use zero-noise extrapolation (ZNE). Specifically, for a given set of parameters, we perform hardware runs at multiple noise factors $G$, and extrapolate the results to the zero-noise limit $G\to 0$ (see Appendix~\ref{app: gate_folding} for details.)

Fig.~\ref{fig: distribution_main} displays the fermion density extracted from the Pauli-$Z$ expectation after ZNE, where we also subtract the contribution from the vacuum to highlight the wave packet's distribution
\begin{equation}\label{eq:delta_ferm_dens}
    \Delta \langle \xi_n^{\dagger} \xi_n \rangle_t = \bra{\psi(t)} \xi_n^{\dagger} \xi_n \ket{\psi(t)} -  \bra{\Omega} \xi_n^{\dagger} \xi_n \ket{\Omega}.
\end{equation}
The fermion density in the vacuum state in the above expression is calculated using the MPS results. 
\begin{figure}[htp!]
    \centering
    \includegraphics[width=0.95\linewidth]{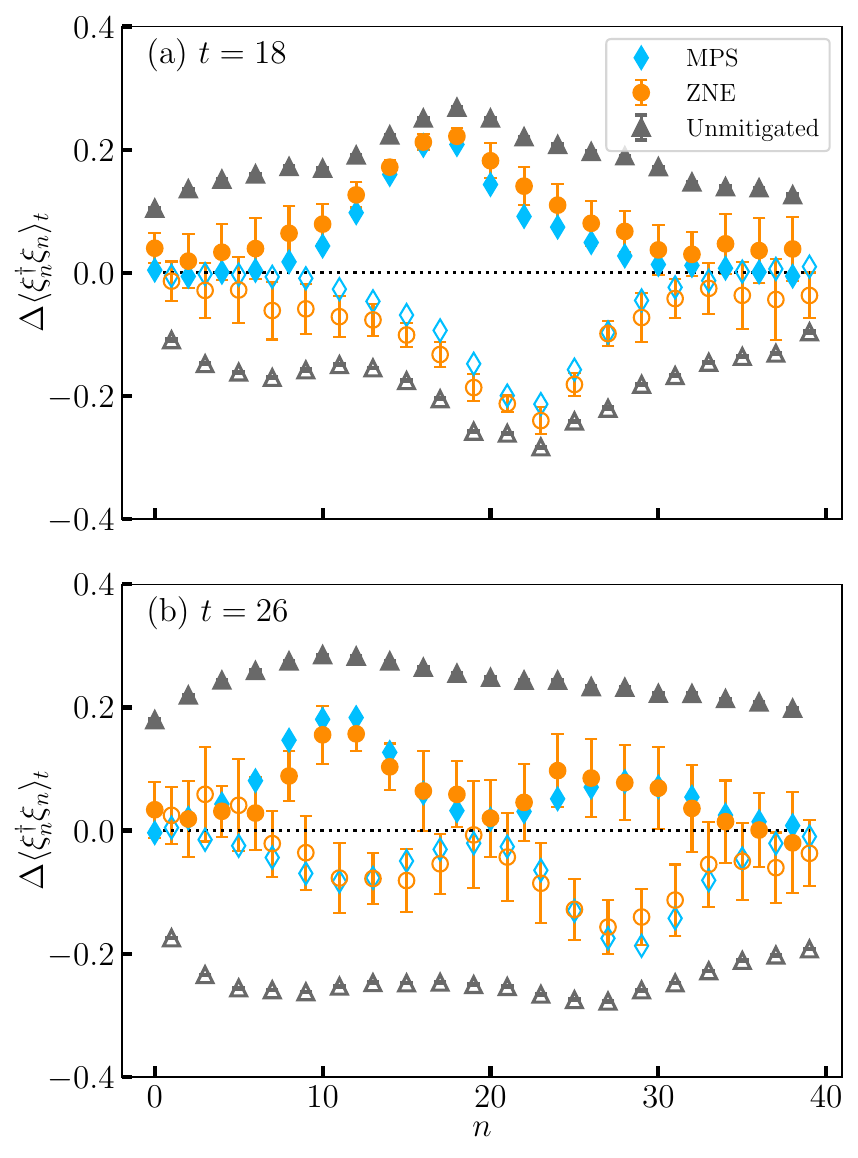}
    \caption{\textbf{Hardware run results for single time slices.} Fermion densities for $(m,g) = (0.4, 0.7)$ at times (a) $t=18$ and (b) $t=26$, comparing quantum hardware runs with precise MPS simulations. Gray triangles show unmitigated hardware results, orange dots show results after ZNE, and blue diamonds represent MPS data. For clarity, markers are filled for even sites and left empty for odd sites. For both panels, the $x$-axis represents the site index, and the $y$-axis represents the fermion density with the vacuum contribution subtracted.
    }
    \label{fig: distribution_main}
\end{figure}
As shown in the figure, the unmitigated hardware results deviate significantly from the ideal MPS simulations, especially at $t=26$, which requires a deeper circuit and more CZ gates. After applying the ZNE, the corrected results agree well with the ideal MPS data across all sites and both time slices.

\begin{figure*}
    \centering
    \includegraphics[width=0.8\linewidth]{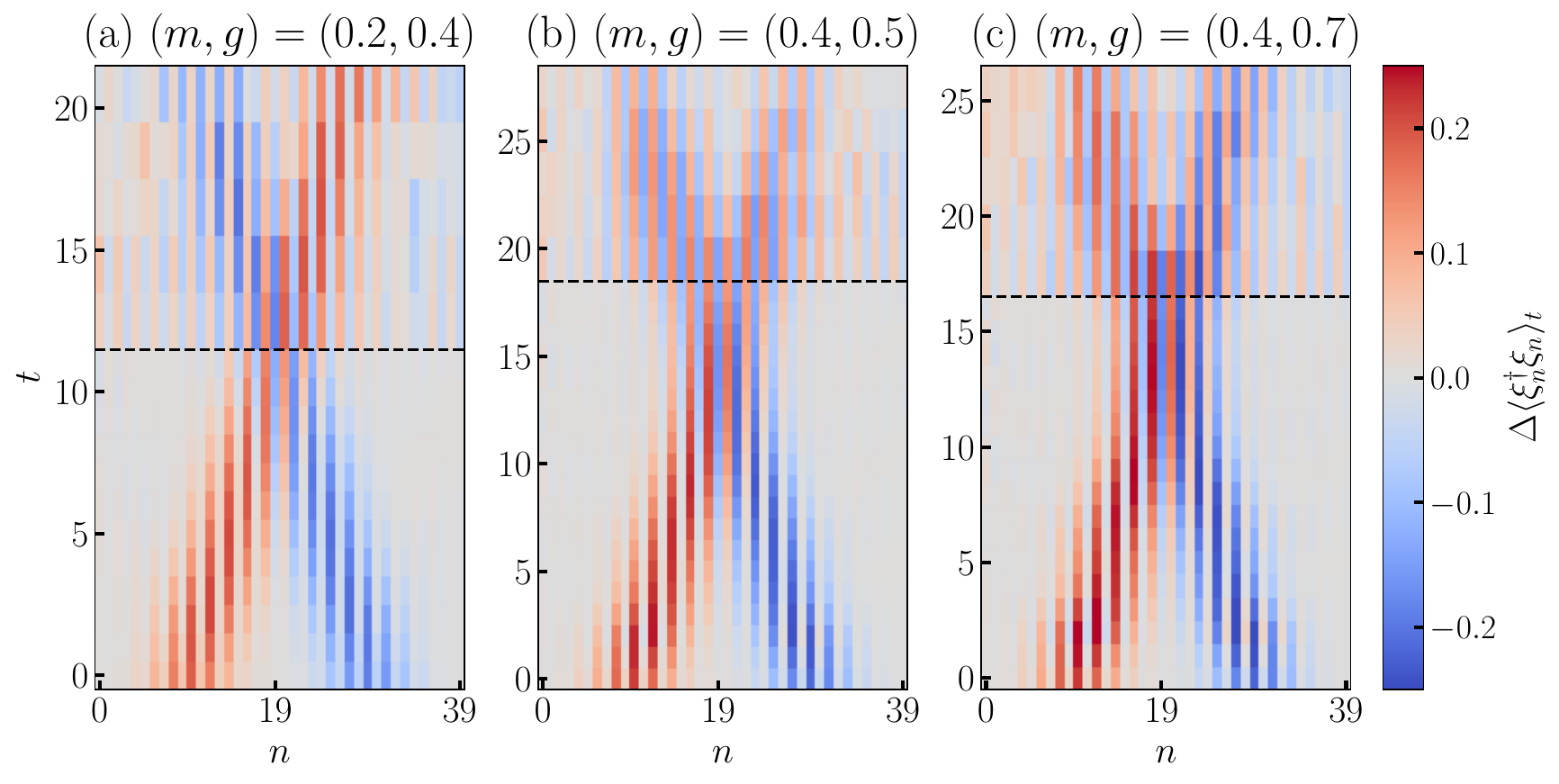}
    \caption{\textbf{Dynamics of the scattering process.} Data for times slices above the dotted horizontal line is obtained from simulations on quantum hardware, while those below are computed using MPS simulations. (a) $(m,g)=(0.2,0.4)$, data are from MPS simulation for $t \leq 11$, and are from hardware run for time $t > 11$. (b) $(m,g) = (0.4, 0.5)$, data are from hardware run for time $t > 18$. (c) $(m,g) = (0.4,0.7)$, data are from hardware run for time $t > 16$. 
    For a comparison with the exact simulation values (computed via tensor network methods) see Fig.~\ref{fig: distribution} in the Appendices).
    }
    \label{fig: dynamics}
\end{figure*}

Finally, the full scattering dynamics for all three cases are shown in Fig.~\ref{fig: dynamics}. 
For time slices before the collision, the results are obtained from the MPS simulation, while for time slices after the collision, the dynamics are executed on the quantum hardware with a timestep of $ t=2$. 
Figure~\ref{fig: dynamics}(a) clearly shows the fermion and antifermion cross each other after $t \sim 11$. In Fig.~\ref{fig: dynamics}(b), the particles have lower velocities due to larger mass, hence the collision takes place at a later time. In addition, each outgoing particle after the collision shows a mixture of the fermion and antifermion components, indicating a partial cross and partial repulsive scattering process. With increased interaction strength, Fig.~\ref{fig: dynamics}(c) shows that most of the fermion and antifermion are reflected after the collision, indicating an elastic scattering in this case. Detailed distributions of the fermion density for each time slice are displayed in Fig.~\ref{fig: distribution} in Appendix~\ref{app: details_40}.

Our results thus show that using tensor network circuit approximation techniques together with a quantum device allow for obtaining a comprehensive picture of the scattering process. The circuit required to prepare the moderately entangled state at an initial time $t_0$ can be efficiently compiled using classical methods. This enables the preparation of the state and its subsequent evolution on a quantum device, thereby entering a regime in which larger amounts of entanglement are generated and classical methods ultimately encounter limitations.

\subsection{Scaling up to larger system sizes: state preparation for 80 qubits}\label{sec: hardware_80}

In order to demonstrate that our approach can also be scaled up to larger system sizes, we benchmark the performance of the  quantum hardware preparing $\ket{\psi(t_0)}$ for a 80-qubit system. Specifically, we consider $(m, g) = (0.2, 0.4)$, and take the MPS-simulated state at $t=10$ as the reference. Similarly to the previous section, this state is then approximated by a parametrized circuit with a two-qubit depth of $24$, using the tensor network circuit approximation approach from Sec.~\ref {sec: variation_state}.  Executing this circuit on hardware allows us to assess the capabilities of larger-scale scattering simulations on current quantum hardware. 

The experiment is carried out on the \texttt{ibm\_fez} device using Qiskit's \texttt{Sampler} primitive. Measurement error mitigation via twirling is employed, along with XY4 dynamical decoupling to suppress decoherence.  
Again, ZNE is employed  to mitigate the effects of noise, where we use the noise factors $G = \{1, 3, 5, 7\}$, corresponding to depths of two-qubit gates $\{25, 73, 121, 169\}$ and total two-qubit gate counts $\{948, 2844, 4740, 6636\}$\footnote{Note that although the two-qubit gate counts for noise factors $G>1$ are exact multiples of the two-qubit gate count for $G=1$, the corresponding two-qubit gate depths generally are not. This discrepancy arises from the partial overlap of certain two-qubit gate structures within the circuit, hence the increase in circuit depth is slightly smaller than the noise factors.}.

\begin{figure}
    \centering
    \includegraphics[width=0.95\linewidth]{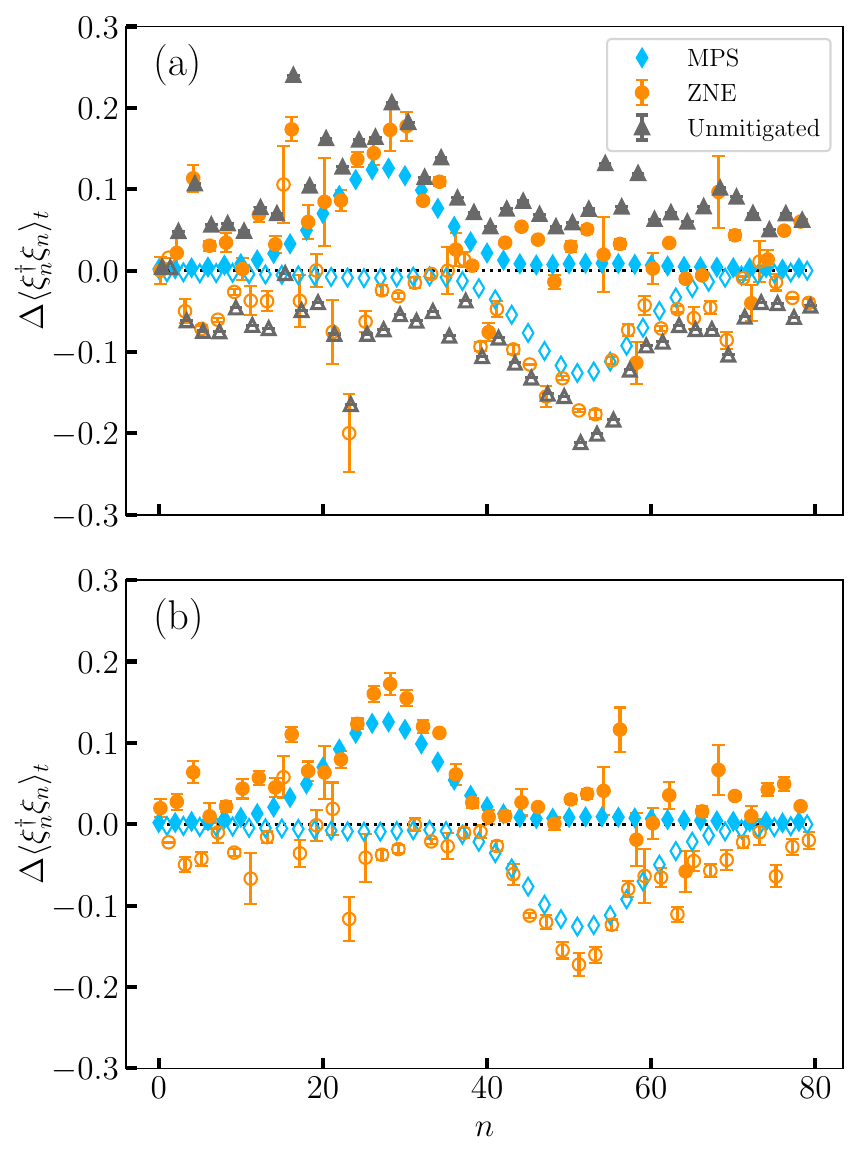}
    \caption{\textbf{Hardware run results for 80 qubit state preparation.} (a) The fermion densities from the MPS simulation (blue diamonds), hardware run result without error mitigation (gray upward triangles), and with ZNE (orange dots). Even sites are shown with filled markers, odd sites with empty markers for visual distinction. (b) Results after ZNE when averaging data related by CP symmetry, which improves outliers and enhances the agreement with the ideal simulation.
    }
    \label{fig:initial_state_80_qubits}
\end{figure}

Figure~\ref{fig:initial_state_80_qubits} compares the results from quantum hardware in comparison with the ideal MPS simulation. 
Even without error mitigation, the raw data from the quantum device qualitatively reproduces the shape of the wave packet. Applying ZNE, just as before, most of the hardware data aligns more closely with the ideal result, as Fig.~\ref{fig:initial_state_80_qubits}(a) shows. However, there are a few data points that show no improvement or even an increased deviation after the ZNE. This is likely the effect of qubits with low performance, as for such large system size we have to use a significant fraction of the qubits on the chip (see Fig.~\ref{fig:fez-layout}). Alternatively, the deep circuit and large number of two-qubit gates for the largest two noise factors might just introduce too much noise, thus rendering making the extrapolation unreliable. 

To improve our data further, we can take advantage of CP symmetry in the model, which in spin language translates to $\langle \sigma_n^z\rangle$ and $\langle \sigma_{N-1-n}^z\rangle$ being equal. Averaging over these two results effectively allows for increasing the statistics and to mitigate outliers. Fig.~\ref{fig:initial_state_80_qubits}(b) presents the ZNE results followed by the CP averaging, where the raw data has been removed to enhance visual clarity. In general, using the averaging procedure, a reduction in the deviation of most of the previous outliers and an overall better agreement with the wave packet distribution is observed. While exploiting the symmetry generally yields an improvement, certain data points still deviate from the ideal results and exhibit large error bars, indicating the need for improved hardware or more advanced error mitigation methods to achieve a similar level of precision as for $N=40$ at this scale.

While we do not perform further time evolution on this 80-qubit state 
due to the prohibitive circuit depths to simulate the full scattering dynamics on available quantum hardware, this benchmark clearly demonstrates the scalability of our tensor network optimized circuit approach for state preparation, as discussed in Sec.~\ref {sec: variation_state}. Future work will investigate real-time dynamics on large systems and improve accuracy, utilizing upgraded hardware and more advanced error mitigation techniques, such as ZNE combined with probabilistic error amplification (PEA).

\section{Summary and Outlook\label{sec:summary_outlook}}
In this work, we demonstrated a hybrid quantum-classical strategy for simulating fermion scattering processes using tensor networks and quantum hardware. By using tensor network optimization to prepare low-entangling states and compress short-time evolution circuits, we significantly reduced the circuit depth compared to conventional methods. This enabled successful hardware runs for 40 qubit dynamics and 80 qubit state preparation on IBM superconducting quantum devices; after error mitigation we achieved results consistent with ideal simulations, even for circuits with two-qubit gate depths up to 96 and total two-qubit gate counts up to 1872.

Specifically, we use the tensor network simulations to prepare the fermion-antifermion scattering state before the collision, where the entanglement is relatively low, and can be constructed by a shallow quantum circuit. This state serves as the initial configuration for execution on quantum hardware. To simulate the post-collision dynamics, where entanglement between subsystems increases due to interactions, we apply a sequence of time-evolution circuits starting from this state, implemented using Trotterized dynamics. Each Trotter step time-evolution circuits can also be variationally optimized and compressed to reduce circuit depth and mitigate hardware noise, without significantly compromising accuracy. By executing these circuits for a sufficiently large number of steps, we are able to probe distinct dynamical behaviors following the interaction. In particular, we observe clear signatures of different scattering outcomes, ranging from transmission to reflection of the particles depending on the initial conditions and the structure of the interactions encoded in the Hamiltonian. 

While classical simulations based on MPS remain feasible for the current system sizes and time scales, they already approach the limits of tractability as entanglement grows with time evolution. In this setting, MPS results can still serve as a valuable benchmark for validating quantum hardware runs. However, we anticipate that near-future experiments, leveraging advanced error mitigation techniques such as probabilistic error cancellation (PEC) and probabilistic error amplification (PEA)~\cite{Temme_2017,Endo2018,Zhang2020,Jinzhao2021}, or post processing tensor network approaches (e.g, TEM)~\cite{Filippov2024} will enable the simulation of increasingly complex scattering phenomena on larger quantum systems  (with 100 qubits and more) and over significantly longer time scales. 

Our method highlights the potential synergies between classical tensor network circuit compression and state propagation on a quantum computer, offering a practical path for simulating scattering in lattice field theories. While our study focuses on the Thirring model, the approach can be extended straightforwardly to other (1+1)D models, such as the Schwinger model. In addition, there are recent developments about the circuit construction for hadronic wave packets~\cite{Davoudi2024, davoudi2025, chai2025}, it would also be an interesting direction to compare the resource cost in these approaches and the tensor network optimization approach. Moreover, incorporating finite lattice spacing and volume effects~\cite{burbano2025_finitevolumn} will be a necessary step toward connecting quantum simulation results to their continuum counterparts. These advances are expected to push quantum simulations beyond the reach of classical methods, opening the door to quantitatively accurate investigations of non-perturbative phenomena in quantum field theory using near-term quantum devices. 

\begin{acknowledgments}
    The authors thank the QC4HEP working group for useful discussions.
    Y.C.\ thanks to Yibin Guo for the helpful discussions about tensor networks.
    J.G. was supported with funding from AWE. ZH acknowledges support from the Sandoz Family Foundation-Monique de Meuron program for Academic Promotion.
    IBM, the IBM logo, and \texttt{ibm.com} are trademarks of International Business Machines Corp., registered in many jurisdictions worldwide. Other product and service names might be trademarks of IBM or other companies. The current list of IBM trademarks
    is available at \url{https://www.ibm.com/legal/} copytrade.
    This work is supported with funds from the Ministry of Science, Research and Culture of the State of Brandenburg within the Center for Quantum Technology and Applications (CQTA). 
    \begin{center}
        \includegraphics[width = 0.08\textwidth]{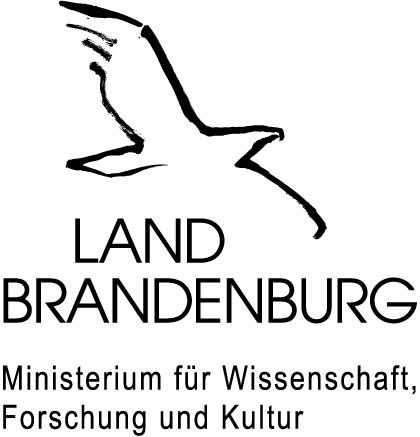}
    \end{center}
\end{acknowledgments}

\appendix

\section{Details on Variational Circuit Compilation}\label{sec:circuit_opt_detail}
In this section we describe the methods used to perform the variational compilation of compact circuits via tensor networks. We perform two circuit optimizations, to generate a circuit to approximately prepare a target state (after application to $|0\rangle$) and to approximate the full unitary of $e^{-iHt}$ for time evolution. 
The methods used largely follow those described in Ref.~\cite{gibbs2024deep}, however we repeat details for clarity.

We use a local optimization based on the Polar Decomposition where, for each $SU(4)$ gate, we compute its environment in the tensor network. Here the gate environment refers to the tensor found when contracting the entire tensor network with the target gate removed. A Polar Decomposition of this tensor finds the optimal 2-qubit unitary to replace the current gate, a standard practice in tensor network optimization~\cite{evenbly2009algorithms}, which has been similarly motivated for circuit optimization~\cite{shirakawa2024automatic, rudolph2023decomposition, gibbs2024deep, causer2024scalable}.

\begin{figure}[t]
    \centering
    \includegraphics[width=0.9\linewidth]{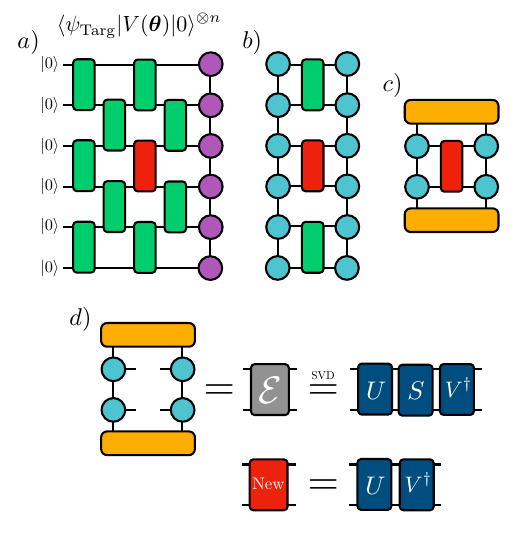}
    \caption{ \textbf{ Illustration: Gate Updates for Variational Compilation} a) Given $|\psi_\text{Targ}\rangle$ represented as an MPS (purple), an illustration of a tensor network representing $\langle \psi_\text{Targ}|V(\boldsymbol{\theta}|0\rangle^{\otimes n}$, where $V(\boldsymbol{\theta})$ is a 1D brickwork circuit. b) For a chosen unitary (red) to be updated, environment MPS (light blue) encode the state of the tensor network outside of the layer. c) To reduce the contraction to tensors local to the chosen unitary, the state of the tensor network due to the outer gates in the layer can be also contracted down to the environment tensors (orange). d) The `gate environment' (grey) of the chosen unitary, denoted $\mathcal{E}$, is then given by the contraction of orange and light blue tensors. The  2-qubit gate is updated by a Polar Decomposition; an SVD of the gate environment finds the optimal unitary to maximize $\big|\langle \psi_\text{Targ}|V(\boldsymbol{\theta})|0\rangle^{\otimes n}\big|^2$  }
    \label{fig: gate_env}
\end{figure}

Computing a gate's environment requires a full contraction of the tensor network representing the inner product required to calculate Eq.~\eqref{eq:C_State} or Eq.~\eqref{eq:C_Uni}. 
An illustration of the environment contraction, and computation of the new unitary is shown in Fig.~\ref{fig: gate_env}.
This is expensive for deep circuits, particularly if it is repeated for every 2-qubit gate in the variational circuit during an optimization sweep. 
We can, however, use an update scheme that exploits the fact that gate updates only cause local changes to the tensor network representing the inner product. 

The primary difference in the circuit optimization for the state preparation and unitary compilation scenarios is the representation of the outer state of the network when computing environments. For state preparation, the state of the tensor network above and below the current layer is represented by environment MPS (by shown in light blue in Fig.~\ref{fig: gate_env}). For the unitary compilation, instead environment MPO are used. The brickwork gate structure and these environment MPS/MPO allowing cheap updates to gates with a layer without recontracting the entire network. Environment tensors are further used to represent the tensor network to the right and left of a particular gate (shown by the orange tensors in Fig.~\ref{fig: gate_env}), allowing cheap sweeps back and forth along the layer.

\begin{figure*}[htp!]         
    \centering
    \includegraphics[width=0.8\linewidth]{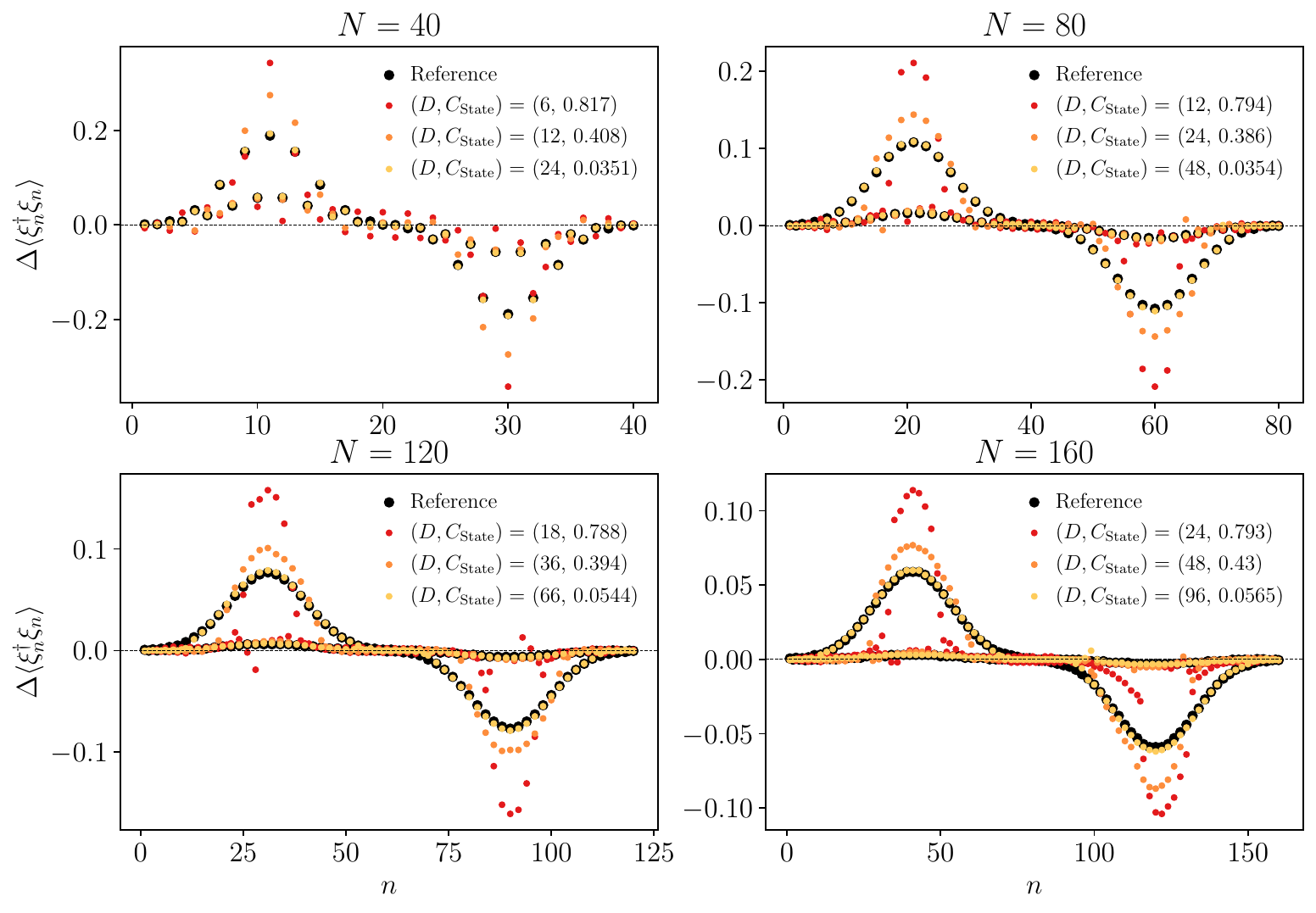} 
    \caption{\textbf{Fermion densities of initial state preparation} Data showing the fermion densities for approximate ground states. We target the initial state preparation described in Section~\ref{sec: variation_state}, for the target state $|\psi_\text{Targ}\rangle = |\psi_0\rangle$ on a scale of system sizes with a number of qubits $N\in\{40,80,120,160\}$. The reference data, shown by black markers, compared against is the high-accuracy MPS simulation of the fermion densities. The colored markers show the fermion densities from the approximate states prepared by MPS-compiled circuits. The circuits are measured in CNOT depth, $D$, where increasing depths result in lower infidelities ($C_\text{State}$), as shown in Fig.~\ref{fig: tn_circuit_mps_vary_n}. 
    }
    \label{fig:fermion_densities_scale_n}
 \end{figure*}
For the state preparation circuit, there is no obvious short-depth circuit to approximate our target state to initialize the optimization with. However, we use an identity initialization, and despite  not incorporating any problem-specific information, the optimization is found to be sufficiently performant to avoid getting stuck in local minima. For the deepest circuits (e.g. circuit depths greater than 0.4 CNOT layers per qubit in Fig.~\ref{fig: tn_circuit_mps_vary_n}), we find benefit from initializing with a previously optimized shorter depth circuit with identity layers appended. For the unitary compression, the 2nd-order Trotter approximation gives a convenient initialization with a depth that can be varied, and this is used throughout our work.

\section{Fermion densities of approximate initial state preparation for increasing system sizes}\label{app:fermion_densities_system_size}
In Fig.~\ref{fig:fermion_densities_scale_n}, we further demonstrate the quality of the variational circuits for scalable initial state preparation shown in Fig.~\ref{fig: tn_circuit_mps_vary_n} by visualizing the delta fermion densities for increasing circuit depth. The reference data of the target state, shown by the black markers, is computed on the system sizes $N \in \{40,80,120,160\}$ using high-accuracy MPS simulation. The colored markers show that for higher fidelity state preparations, given by lower values of the cost function $C_\text{State}$, for deeper circuits does indeed show rapid convergence in the delta fermion densities, with $C_\text{State} \approx 0.05$ closely reproducing the reference data. 

\section{Resource estimation for the scattering process using the convention approach}\label{app: resources}
In this appendix, we analyze the circuit depth and two-qubit gate requirements for realizing the full scattering process using the conventional approach, i.e., decomposing the wave packet operators using the Givens rotation~\cite{Chai2025fermionicwavepacket} and approximating the time evolution by second-order Trotterization.

First, the vacuum state $|\Omega\rangle$ must be prepared. This can be achieved using a variational quantum eigensolver (VQE) with a few layers of parameterized iSwap gates, as discussed in Appendix D of Ref.~\cite{Chai2025fermionicwavepacket}. Since this can usually be achieved by a shallow circuit, we neglect this part in resource estimation for simplicity.

\begin{figure*}
    \centering
    \includegraphics[width=0.8\linewidth]{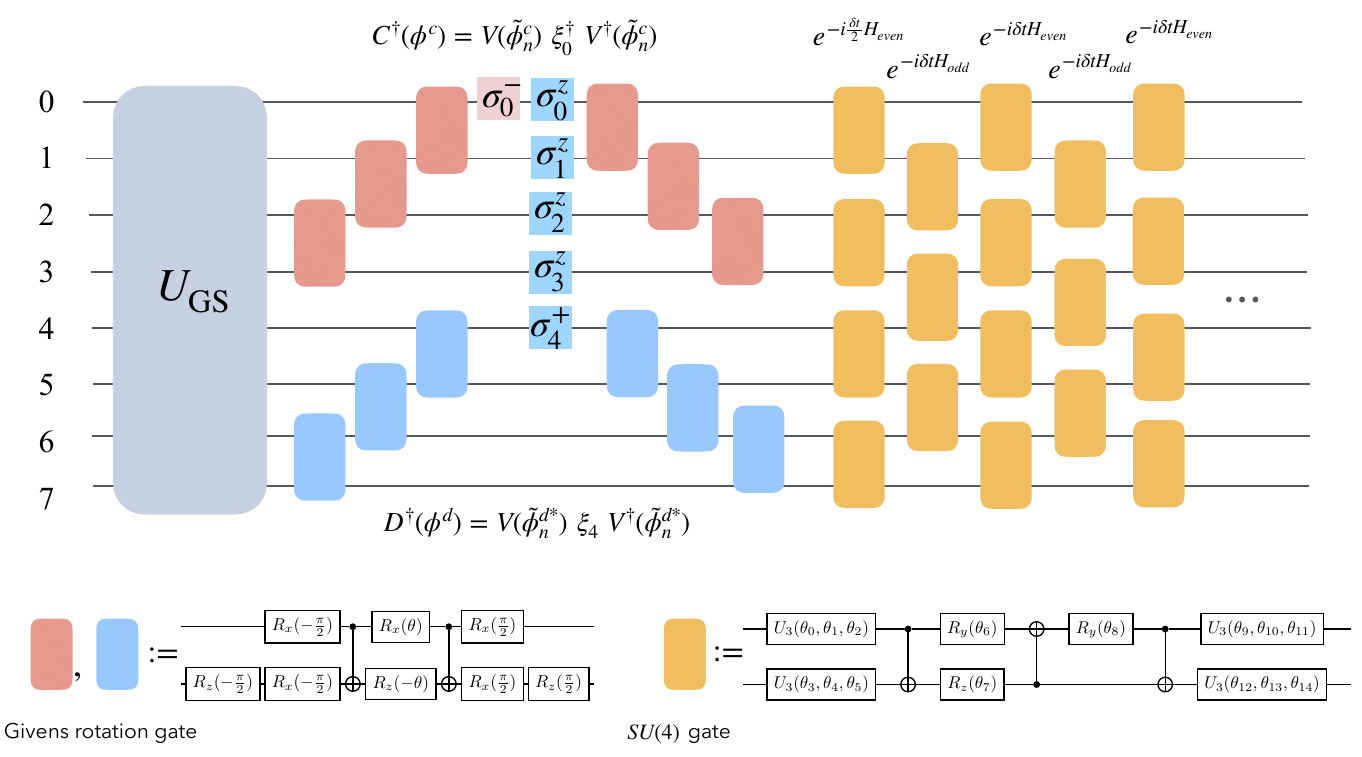}
    \caption{\textbf{Illustration of the circuit for simulating scattering using the conventional approach.} The example shows an 8-qubit system,  where the gray box represents the VQE ansatz $U_\text{GS}$ to prepare the vacuum state. The following red and blue boxes represent the gates implementing the Givens rotation to create fermion and antifermion wave packets, $U_\text{WP}$. The fermionic operators $\xi_0^{\dagger}, \xi_4$ are transformed to Pauli operators with the Jordan-Wigner transformation in Eq.~\eqref{eq:jw}. The yellow boxes corresponds to the Trotterized time evolution circuit, where each layer consists of $SU(4)$ gates.}
    \label{fig:conventional_circuit}
\end{figure*}

Once the vacuum is prepared, the initial state of scattering $|\psi_0\rangle = D^{\dagger} C^{\dagger} |\Omega\rangle$ is constructed by applying fermion and antifermion wave packet creation operators. As introduced in Ref.~\cite{Chai2025fermionicwavepacket}, the operators $C^{\dagger}, D^{\dagger}$ can be realized by unitary operators $V(\tilde{\phi}_n^c), V(\tilde{\phi}_n^{d\ast})$, each of the unitary can be decomposed to $(N/2-1)$ two-qubit operators $V(r)$ using Givens rotations as detailed in Eq.~(17-19) of Ref.~\cite{Chai2025fermionicwavepacket}.
\begin{align}
    \begin{aligned}
        C^{\dagger}(\phi^c) & \sim \sum_{n=0}^{N/2-1} \xi_n^{\dagger} \tilde{\phi}_n^c  = V(\tilde{\phi}_n^c)  \xi_0^{\dagger}  V^{\dagger}(\tilde{\phi}_n^c), \\
        D^{\dagger}(\phi^d) & \sim \sum_{n=N/2}^{N-1} \xi_n \tilde{\phi}_n^d  = V(\tilde{\phi}_n^{d\ast})  \xi_{N/2}  V^{\dagger}(\tilde{\phi}_n^{d\ast}),
    \end{aligned}
    \label{eq: givens_fermion_op}
\end{align}
Here use a simplification compared with Eq.~\eqref{eq:creation_operators_gaussian_particle}: the fermionic wave packet is defined only on the first half of the lattice with its tail truncated beyond the midpoint, and the antifermionic wave packet is similarly localized to the second half.  This gives a good approximation of Eq.~\eqref{eq:creation_operators_gaussian_particle} since the wave packets are localized and the system size $N=40$ in this work is large enough to ensure the two wave packets are well separated. Thus, each wave packet involves $2 \times (N/2 - 1)$ local operators $V(r)$ arranged linearly, where the factor of 2 accounts for the presence of both $V(\phi_n)$ and $V^{\dagger}(\phi_n)$ in each packet. Each of $V(r)$ can be decomposed by two-CNOT gates as shown in Fig.~\ref{fig:conventional_circuit}. The operator $\xi_0^{\dagger}, \xi_{N/2}$ is not unitary and needs a Hadamard test circuit to realize it, introducing one ancilla qubit and one CNOT gate for each wave packet; we omit this contribution in the final gate counts for simplicity. Since the two wave packet circuits can be executed in parallel as shown in Fig.\ref{fig:conventional_circuit}, the CNOT depth for this step is $2N-4$, with a total of $4N-8$ CNOT gates.

About the time evolution operator $U(t) = e^{-iHt}$, we employ the second-order Trotterization with timestep $\Delta t$
\begin{equation}
    \begin{aligned}
            U(t) &\approx \left( e^{-i\frac{\Delta t}{2} H_\text{even}} ~ e^{-i\Delta t H_\text{odd}} ~ e^{-i\frac{\Delta t}{2} H_\text{even}} \right)^{\frac{t}{\Delta t}}, \\
    \end{aligned}\label{eq:trotter}
\end{equation}
with,
\begin{equation}
    \begin{aligned}
        H_{\mathrm{even}} &= \frac{1}{4} \sum_{n\in 0, 2 , 4,\dots}^{N-2} \left( \sigma_{n+1}^y\sigma_{n}^x - \sigma_{n+1}^x\sigma_{n}^y \right),\\
        &+ \frac{m}{2}\sum_{n\in 0, 2 , 4,\dots}^{N-1} (-1)^n(1-\sigma^z_n)\\
        &+ \frac{g}{4}\sum_{n\in 0, 2 , 4,\dots}^{N-2} (I-\sigma_{n+1}^z ) (I - \sigma_{n}^z).
    \end{aligned}
\end{equation}
$H_{\mathrm{odd}}$ has the similar form but defined at odd sites. In the Eq.\eqref{eq:trotter}, the first term $e^{-i\Delta t H_\text{even}/2}$ in timestep $k$ can be compressed with the third one from timestep $k-1$, resulting in reduced circuit depth. Each of the term $e^{-i\Delta t  H_{\mathrm{even(odd)}}}$ corresponds to $N/2 \text{ } (\text{or } N/2-1)$ parallel $SU(4)$ gates as shown Fig.~\ref{fig:conventional_circuit}. Each $SU(4)$ gate can be decomposed into at most 3 CNOT gates, so one more full Trotter step adds a depth of $6$, and $ 3(N-1)$ CNOT gates. The first timestep needs one more $SU(4)$ layers as in Eq.~\eqref{eq:trotter}.

Finally, we give the resource estimation for the entire scattering process with total time $T$ and system size $N$ in the Table.~\ref{tab:resource_trotter}. For the time evolution, the additional 3 CNOT layer and $(3/2)N$ CNOT gates come from the first term $e^{-i\Delta tH_{\mathrm{even}}/2}$ of the first timestep in Eq.~\eqref{eq:trotter}.
\begin{table}[htp!]
    \centering
    \begin{tabular}{@{} lcc @{}}
        \toprule
                    & CNOT layers & CNOT gates \\
        \midrule
        Wave packet    & $2N-4$   & $4N-8$   \\
        \midrule
        Time evolution & $6 T/\Delta t + 3$   & $3 (N-1) T/\Delta t + (3/2) N$\\
        \bottomrule
    \end{tabular}
    \caption{Resource estimation for the full scattering process using Givens rotation to prepare wave packets and Trotterization for time evolution.}
    \label{tab:resource_trotter}
\end{table}

Take a 40-qubit system as an example, preparing the wave packet requires a circuit depth of 76 and 152 CNOT gates. The subsequent time evolution circuit with $T=28, \Delta t = 2/3$ add another 255 layers and 4974 two-qubit gates. This highlights the substantial resource cost of simulating long-time dynamics to capture the full scattering process. Although these resource estimates are not prohibitive in principle, they present a significant challenge for current devices, particularly when aiming for accurate, high-fidelity results.

\section{Zero-noise extrapolation}\label{app: gate_folding}
In this work we use gate folding to realize different noise levels to be able to carry out ZNE. To realize digital gate folding~\cite{Temme_2017, Li_2017, Kandala_2019, Giurgica_Tiron_2020, Pascuzzi_2022, Kim_2023} we use the functionality in Qiskit Runtime~\cite{qiskit2024} to amplify noise. Specifically, it replaces two-qubit gates $U$ is by $U(UU^{\dagger})$, which yields the same logical operation but increases the circuit depth, effectively leading to a noise amplification by factor of $3$ for corresponding two-qubit gate. Higher noise factors of $5$ and more can be obtained similarly by the replacing the gate by the sequence $U(UU^{\dagger})(UU^{\dagger})$. Applying this globally to every two-qubit gate in the circuit allows for directly realizing odd noise factors $G=3,5,7$. Even values of $G$ or fractional values can realized by applying this technique to a fraction of two-qubit gates in the circuit, which are chosen randomly. A value of $G=1$ corresponds to the original circuit.

\begin{figure}[htp!]
    \centering
    \includegraphics[width=0.95\linewidth]{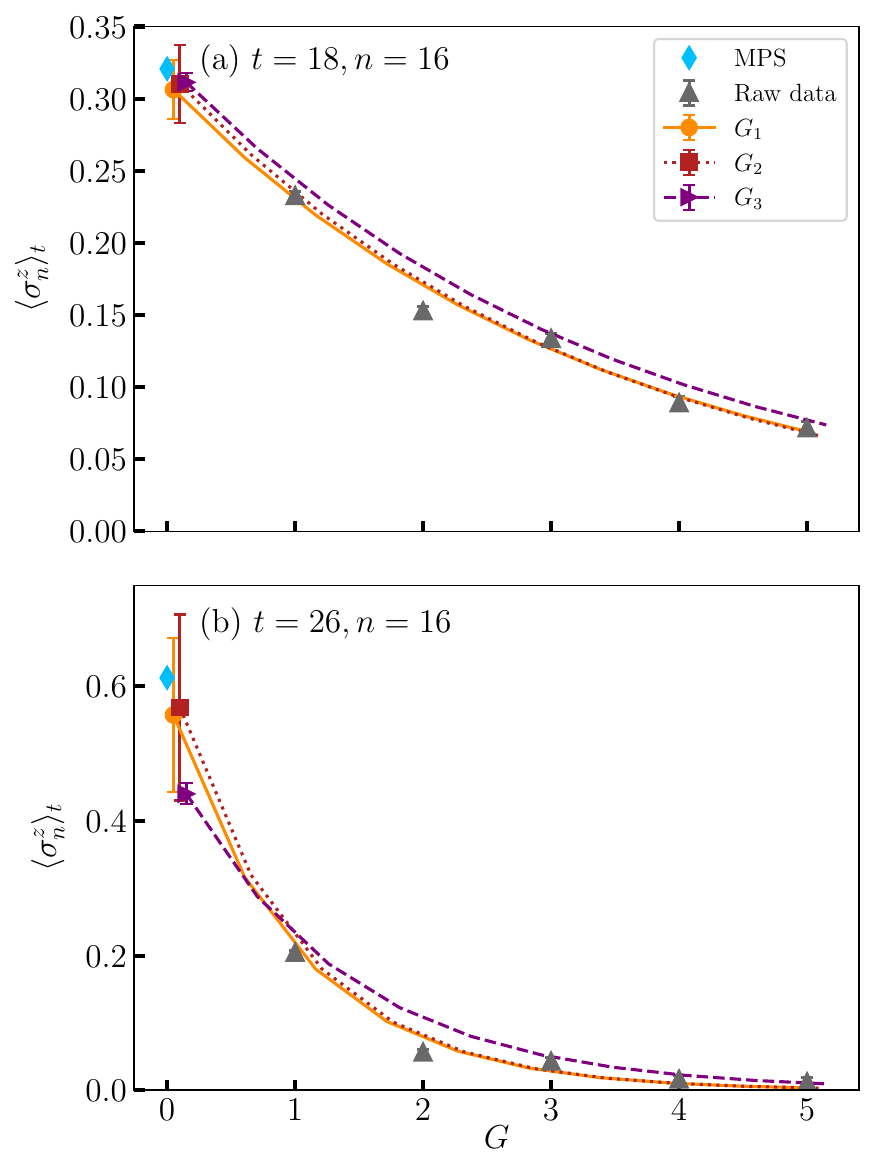}
    \caption{\textbf{Extrapolation of hardware run results.} (a) Results for $(m,g) = (0.4, 0.7)$ at times $t=18$, and (b) $t=26$. The expectation value $\langle \sigma_n^z \rangle$ at site $n=16$ is shown at different noise factors $G$. Blue diamonds correspond to the noiseless results from a precise MPS simulation. Gray triangles show the raw quantum hardware data at different noise factors, with only error suppression techniques (dynamical decoupling, twirled readout error extinction, and Pauli twirling) applied. The three different lines represent an exponential fit to the hardware data using different subsets of noise factors, $G_1 = \{1,2,3,4,5\}$, $G_2 = \{1,2,3,4\}$ and $G_3 = \{1,3,5\}$, the fitted noiseless results are shown at $G=0$, where the different data points are offset slightly in horizontal direction for better visibility.}
    \label{fig: ZNE_extrapolation}
\end{figure}

In Fig.~\ref{fig: ZNE_extrapolation}, we present an example for the ZNE for the data presented in Sec.~\ref{sec: hardware_40} the quantum hardware $m=0.4, g=0.7$ for two time slices corresponding to the shallowest ($t=18$) and deepest ($t=26$) circuits that we executed respectively.
Specifically, the first row of Fig.~\ref{fig: ZNE_extrapolation} illustrates the ZNE for the Pauli-$Z$ expectation value for site $n=16$ for time slices $t=18$ (panel (a)) and $t=26$ (panel (b)). For the deeper circuit, corresponding to $t=26$, we observe that for our largest noise factor, $G=5$, the expected value approaches zero, thus indicating that we have reached a noise level where the state is fully depolarized. Furthermore, it can be observed that even noise factors, and especially $G=2$, influence the results proportionally more than odd noise factors. To probe for effects of these observations, we perform an exponential extrapolation using different subsets of noise factors as outlined in the caption. In particular, we try discarding the largest noise factors and the even values of $G$. While for $t=18$ all fits yield a central value compatible with the exact result from the MPS simulation, we observe that for $t=26$ the extrapolation only using odd values $G$ systematically underestimates the result. In addition, compared to the results for $t=18$, the central value for $t=26$ has a larger error bar reflecting larger uncertainties due to more noise in deeper circuit. Since for both cases including all available noise factors in the fit yields consistent results, we use use this method for all the data shown in the following.
\begin{figure*}
    \centering
    \includegraphics[width=0.88\linewidth]{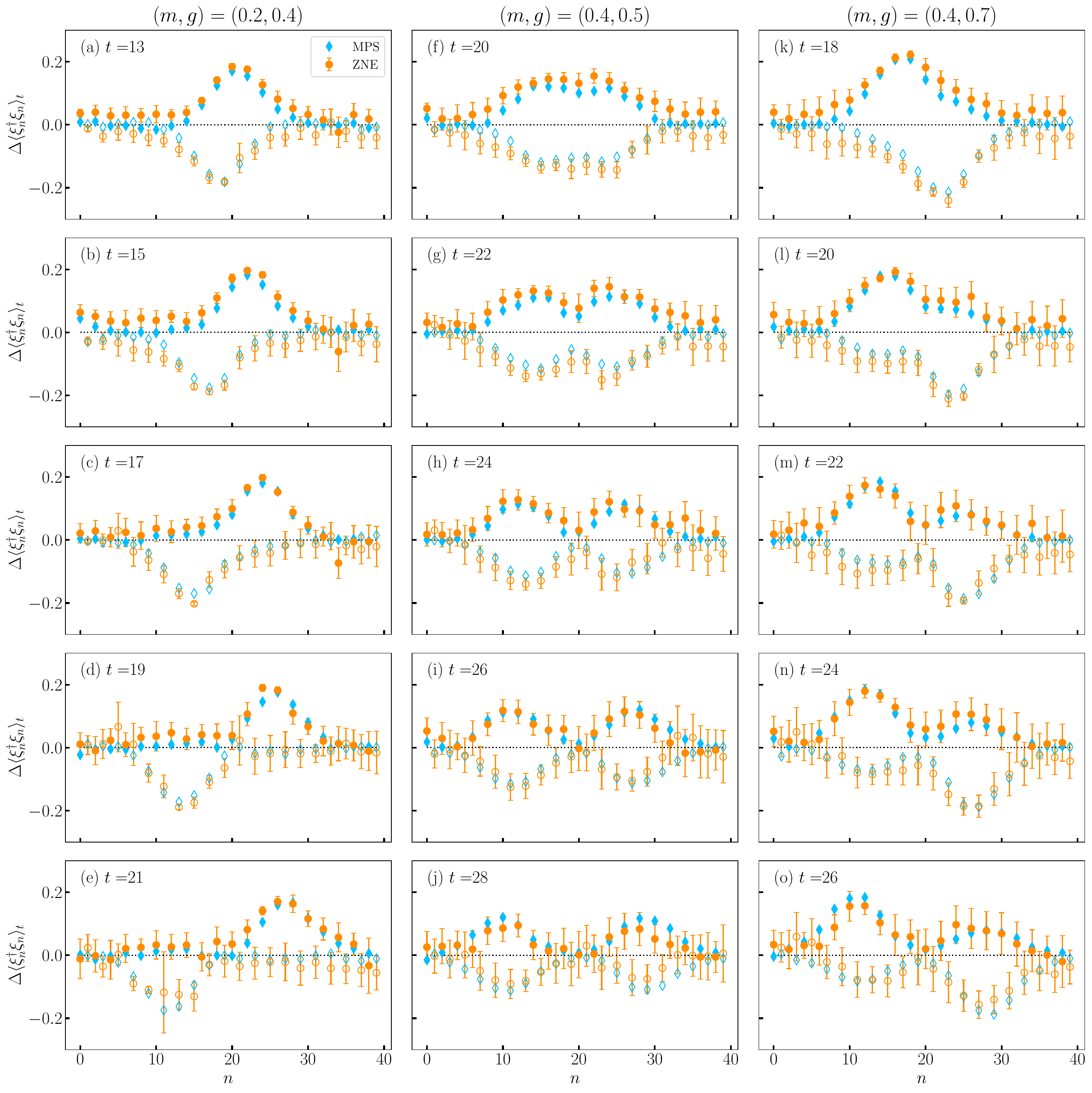}
    \caption{\textbf{Distribution of particle densities for time slices executed on hardware.} Different columns correspond to (a)-(e) $(m,g) = (0.2, 0.4)$; (f)-(j) $(m,g) = (0.4,0.5)$; and (k)-(o) $(m,g)= (0.4,0.7)$. Blue diamonds correspond to the results from MPS simulations, and orange dots represent the hardware run result with ZNE, where error bars include both the statistical uncertainties due to finite measurements and the extrapolation effect. For clarity, markers are filled for even sites and left empty for odd sites. As a visual guide, the horizontal dashed line indicates the zero value of $\Delta \langle\xi_n^{\dagger}\xi_n \rangle$.}
    \label{fig: distribution}
\end{figure*}

\section{Details of hardware run for 40 qubits}\label{app: details_40}
In this appendix, we present the detailed fermion density distribution after collision in 40-qubit systems, compare results from hardware runs and MPS simulations. For all three cases, $(m,g) \in \{(0.2, 0.4), (0.4, 0.5), (0.4, 0.7)\}$, the hardware results are quite consistent with the ideal results at the smallest timestep (the first row in Fig.~\ref{fig: distribution}). As time increases, deeper circuits are required, leading to larger statistical fluctuations and less stable extrapolated results, as reflected by the increasing error bars — particularly in the longest time case (the bottom row of Fig.~\ref{fig: distribution}). Nevertheless, even for circuits with depths up to 96 layers, as shown in Fig.~\ref{fig: distribution}(j) and (o), the mitigated hardware results successfully capture the expected physical behavior.

\FloatBarrier
\bibliography{references}
\end{document}